\documentstyle[a4,12pt,epsf,amsfonts,amssymb]{article} 

\DeclareFontFamily{U}{rsfs}{\skewchar\font"7F}
\DeclareFontShape{U}{rsfs}{m}{n}{
	<-6> rsfs5
	<6-8> rsfs7
	<8-> rsfs10
	}{}
\DeclareMathAlphabet{\mathscr}{U}{rsfs}{m}{n}

\def\bena{\begin{eqnarray}}
\def\eena{\end{eqnarray}}
\def\f(#1/#2){\frac{#1}{#2}} 
\def\scalar{{\Bbb S}} 
\def\vector{{\Bbb V}} 
\def\tensor{{\Bbb T}}

\newcommand{\OD}{\widehat \nabla} 
\newcommand{\BD}{D} 
 
\newcommand{\K}{{\mathscr K}} 

\newcommand{\re}{{\rm Re}}

\newcommand{\non}{\nonumber}

\begin{document}  

\title{Dynamics in Non-Globally-Hyperbolic Static Spacetimes III: 
       Anti-de Sitter Spacetime} 

\author{Akihiro Ishibashi$^{\dag}$ and Robert M. Wald$^{\ddag}$ \\ \\ 
 {\it $^{\dag}$Department of Applied Mathematics and Theoretical Physics,} \\ 
 {\it Centre for Mathematical Sciences,} \\
 {\it University of Cambridge, Wilberforce Road,}\\ 
 {\it Cambridge CB3 0WA, UK} \\ \\
 {\it $^{\ddag}$Enrico Fermi Institute and Department of Physics}\\ 
 {\it University of Chicago}\\ 
 {\it 5640 S. Ellis Avenue}\\ 
 {\it Chicago, Illinois 60637-1433, USA} 
         } 
\maketitle 

\begin{abstract} 
In recent years, there has been considerable interest in theories
formulated in anti-de Sitter (AdS) spacetime. However, AdS spacetime
fails to be globally hyperbolic, so a classical field satisfying a
hyperbolic wave equation on AdS spacetime need not have a well
defined dynamics. Nevertheless, AdS spacetime is static, so the
possible rules of dynamics for a field satisfying a linear wave
equation are constrained by our previous general analysis---given in
paper II---where it was shown that the possible choices of dynamics
correspond to choices of positive, self-adjoint extensions of a
certain differential operator, $A$. In the present paper, we reduce
the analysis of electromagnetic, and gravitational perturbations in
AdS spacetime to scalar wave equations. We then apply our general
results to analyse the possible dynamics of scalar, electromagnetic,
and gravitational perturbations in AdS spacetime.  In AdS spacetime,
the freedom (if any) in choosing self-adjoint extensions of $A$
corresponds to the freedom (if any) in choosing suitable boundary
conditions at infinity, so our analysis determines all of the possible
boundary conditions that can be imposed at infinity. In particular, we
show that other boundary conditions besides the Dirichlet and Neumann
conditions may be possible, depending on the value of the effective
mass for scalar field perturbations, and depending on the number
of spacetime dimensions and type of mode for electromagnetic and
gravitational perturbations.
\end{abstract}


\section{Introduction} 
\label{sect:1}

Anti-de Sitter (AdS) spacetime\footnote{
In this paper, by $m$-dimensional AdS spacetime of curvature
radius $\ell$, we mean the universal covering space of the hyperboloid
composed of points whose squared distance from the origin is $-\ell^2$ in
${\Bbb R}^{m+1}$ with a flat metric of signature $--+...+$. 
} 
has come to play a central role in a
number of contexts in theoretical physics~\cite{Gibbons2001}, most
prominently in the ``AdS-CFT'' correspondence~\cite{Mal,GKP,Witten}
(see also e.g., \cite{AGMOO} and references therein). Although AdS
spacetime is maximally symmetric and geodesically complete, it admits
no Cauchy surface and thus provides a very simple but nontrivial
example of a non-globally-hyperbolic spacetime. Consequently, it is
far from obvious, a priori, that one can define a sensible, 
deterministic field dynamics in AdS spacetime, even for classical, 
linear fields: If a classical field satisfies a linear hyperbolic 
equation, then it is locally uniquely determined by suitable initial 
data on a spacelike hypersurface; however, in order to guarantee that 
a solution corresponding to given initial data exists globally and 
is globally unique, it is essential that the spacetime be globally 
hyperbolic and that the data be specified on a Cauchy surface (see, 
e.g.~\cite{Wald1984}).

For the case of a scalar wave equation in AdS spacetime, it is clear
that---at least for some values of mass and coupling to curvature---it
is necessary to specify boundary conditions at infinity in order to
obtain a well posed, deterministic evolution law. This issue has been
investigated by a number of authors (see, in particular,
\cite{AIS1978,BF1,BF2}) and some proposals---such as Dirichlet or
Neumann conditions---have been made. However, it is far from obvious
that the boundary conditions that have been proposed encompass all the
possibilities for dynamical evolution laws in AdS spacetime. New
dynamical evolution laws would provide new ``bulk theories'' in AdS
spacetime, and could be of interest in view of the AdS-CFT
correspondence\footnote{ 
Some implications of different choices of boundary conditions for 
AdS-CFT correspondence have been discussed in~\cite{MR2000,MR2001}. 
In particular, it was pointed out that multi-trace deformation 
of a boundary quantum field theory can be incorporated in the AdS-CFT 
correspondence by using a general boundary condition for the  
corresponding bulk field~\cite{Witten2001,BSS2001} 
(see also~\cite{Minces2003,Minces2004} and references therein). 
}, among other reasons. 

The purpose of this paper is to apply our general analysis of dynamics
in static, non-globally-hyperbolic spacetimes~\cite{IW2003a} to
systematically determine all possible boundary conditions that can be
imposed at infinity for scalar, electromagnetic, and gravitational
perturbations of AdS spacetime. In our previous paper~\cite{IW2003a},
we proved that any dynamical evolution law for a scalar field 
in a static, stably causal spacetime that (i)
locally agrees with the wave equation, (ii) admits a suitable
conserved positive energy, (iii) is compatible with the time
translation and time reflection symmetries, and (iv) satisfies a
certain convergence condition must correspond to a prescription of the
type first proposed in~\cite{Wald1980}. Consequently, the general
prescription for a dynamical evolution law for a scalar field in a
static, stably causal spacetime arises as follows:

Let $(M,g_{\mu \nu})$ be a stably causal (but not necessarily globally
hyperbolic) spacetime which admits a hypersurface orthogonal timelike
Killing vector field, $t^\mu$, whose orbits are complete. Let $\phi$
be a scalar field on $(M,g_{\mu \nu})$ which satisfies the
Klein-Gordon equation
\bena 
\nabla^\mu \nabla_\mu \phi - m_0^2 \phi = 0 \,,
\label{eq:Klein-Gordon}
\eena 
with $\nabla_\mu$ being the derivative operator associated with 
the spacetime metric $g_{\mu \nu}$. 
Since $(M,g_{\mu \nu})$ is static, the equation of 
motion~(\ref{eq:Klein-Gordon}) can be rewritten as 
\bena 
\label{A}
   \frac{ \partial^2}{ \partial t^2} \phi = - A \phi \,,  
\eena  
where $t$ denotes the Killing parameter and 
\bena 
  A = - VD^i (VD_i) +  m_0^2 V^2 \,, 
\label{ope:differential} 
\eena 
where $V\equiv (-t^\mu t_\mu)^{1/2}$ and $D_i$ is the derivative
operator compatible with the induced metric defined on a spacelike
hypersurface $\Sigma$ orthogonal to the orbits of $t^\mu$.  

We can view $A$ as an operator (with domain $C^\infty_0(\Sigma)$) on
the Hilbert space, $L^2(\Sigma,V^{-1}d\Sigma)$, of square integrable
functions on $\Sigma$ with volume measure $V^{-1} d\Sigma$, where
$d\Sigma$ denotes the natural volume element on $\Sigma$ arising
from the induced metric.  Then,
provided that $m_0^2 \geqslant 0$, $A$ is a positive, symmetric operator
on $L^2(\Sigma,V^{-1}d\Sigma)$. Hence, $A$ always admits at least one
positive self-adjoint extension---namely, the Friedrichs extension,
$A_F$.  Let $A_E$ denote any positive self-adjoint extension of
$A$. For any initial data $(\phi_0,\dot \phi_0) \in
L^2(\Sigma,V^{-1}d\Sigma) \times L^2(\Sigma,V^{-1}d\Sigma)$, define
$\phi_t$ by
\bena 
\phi_t = \cos(A_E^{1/2}
t) \phi_0 + A_E^{-1/2} \sin(A_E^{1/2} t) \dot{\phi}_0 \,.
\label{def:dynamics} 
\eena 
Then it was shown in \cite{Wald1980} that whenever $(\phi_0,\dot
\phi_0) \in C^\infty_0(\Sigma) \times C^\infty_0(\Sigma)$, the
quantity $\phi_t$ defines a smooth solution of
eq.~(\ref{eq:Klein-Gordon}) on all of $M$. Furthermore, this solution
agrees with the solution determined by the initial data $(\phi_0,\dot
\phi_0)$ within the domain of dependence of initial hypersurface
$\Sigma$. In addition, these solutions have a conserved, positive
energy $E = (\phi_t, A_E \phi_t)_{L^2} + (\dot{\phi}_t,
\dot{\phi}_t)_{L^2}$, and satisfy properties (iii) and (iv) mentioned
above.

As already stated, the main result proven in \cite{IW2003a} was that the
above prescription eq.~(\ref{def:dynamics}) is the only possible way
for defining dynamics, subject to requirements (i)-(iv). Hence, we
have the following simple algorithm to systematically determine the
possible boundary conditions at infinity for a scalar field in AdS
spacetime: (a) Write the equation of motion in the form (\ref{A}),
thereby identifying the operator $A$,
eq.~(\ref{ope:differential}). (b) Determine all of the positive, 
self-adjoint extensions, $A_E$, of $A$. (c) Interpret the dynamics 
defined by eq.~(\ref{def:dynamics}) in terms of boundary conditions at
infinity. In the present paper, we shall carry out this procedure to
completion.

The only property of $A$ used in the above analysis is that it is a
positive, symmetric, elliptic operator on a suitable $L^2$ Hilbert
space. Therefore, the analysis of scalar field dynamics 
given in~\cite{Wald1980} and~\cite{IW2003a} can be generalised 
straightforwardly to second order, linear, hyperbolic equations
satisfied by arbitrary tensor fields provided that in static 
spacetimes\footnote{The prescription of~\cite{Wald1980} has recently
been generalised to stationary spacetimes~\cite{Itai2003}. However, no
uniqueness results similar to those of~\cite{IW2003a} are presently
known for the stationary case.} these equations can still be put in
the general form of eq.~(\ref{A}), with $A$ a positive, symmetric,
elliptic operator. In particular, it can be applied to Maxwell's
equations and the linearised Einstein equation provided that these
equations have been put in the form eq.~(\ref{A}). However, it is not
straightforward to put Maxwell's equations or the linearised Einstein
equation in the required form. For example, although Maxwell's
equations for the vector potential, $A_\mu$, in the Lorentz gauge have
the desired second order, hyperbolic form, the ``inner product''
needed to make the corresponding $A$ symmetric is not positive
definite, since it yields a negative contribution from the time
component, $A_0$, of $A_\mu$. 
In Minkowski spacetime, one may set
$A_0=0$ by a further gauge choice, but such a choice is not possible
in a general curved spacetime. Similar remarks apply to the linearised
Einstein equation in the transverse traceless gauge.

In this paper, we shall analyse Maxwell's equations and the linearised
Einstein equation in AdS spacetime by making use of the spherical
symmetry of AdS spacetime to decompose the vector potential and metric
perturbation into their ``scalar'', ``vector'', and ``tensor'' parts
with respect to the rotation group. We will then expand each part
in the appropriate spherical harmonics and will reduce the resulting
equations to a collection of decoupled scalar wave equations in two
dimensions, each of which is of the general form (\ref{A}). In fact,
we will show that each spherical harmonic component of the scalar wave
equation (with arbitrary mass and curvature coupling), Maxwell's
equation, and the linearised Einstein equation {\it all take exactly
the same general form}, the only difference being the values of two
parameters that appear in this equation. Therefore, we may determine
the possible boundary conditions at infinity for scalar,
electromagnetic, and gravitational perturbations of AdS spacetime by
studying the self-adjoint extensions of a two parameter family of
second order differential operators in one variable.

In our analysis, we shall consider $(n+2)$-dimensional AdS spacetime
$(M, g_{\mu \nu})$
for all $n\geqslant 1$. (Our results also can be straightforwardly extended
to $n=0$ for the case of a scalar field.) We will work in a 
globally defined coordinate
system in which the metric takes the form
\bena
 ds^2_{(n+2)} = \frac{\ell^2}{\sin^2x}
                \left(
                      -dt^2 + dx^2 + \cos^2 x \gamma_{ij}dz^idz^j
                \right) \,, 
\label{metric:ads} 
\eena 
where $\gamma_{ij}(z)dz^idz^j$ is the metric of the $n$-dimensional round
unit sphere $S^n$ and $\ell$ denotes the curvature radius.  The range
of $x$ is $(0,\pi/2]$ and the conformal boundary is located at $x =
0$. The coordinate vector field $\partial/\partial t$ is a globally
timelike, hypersurface orthogonal Killing vector field\footnote{ 
The analysis of dynamics in this paper will be performed with respect 
to this globally timelike Killing field. There are other hypersurface 
orthogonal Killing fields in AdS spacetime that are timelike in 
a region that includes at least a portion of infinity.  
We believe that an analysis of dynamics with respect to these 
static Killing fields would give rise to precisely the same choices 
of boundary conditions, but we have not fully investigated this
issue.

There have arisen some interests in implications of the topology 
for the AdS-CFT correspondence~\cite{WY1999,GSWW,MM2004}.   
The simplest situation can be obtained by considering a quotient 
of AdS spacetime by a certain discrete subgroup of the group of 
the AdS isometries. We believe that an analysis of dynamics in 
such a geometry parallels the present analysis performed 
in the universal covering space. 
}. 
By use of the spherical harmonic decomposition described in the
previous paragraph, we will reduce the scalar wave equation, Maxwell's 
equation, and the linearised Einstein equation to wave equations 
in a $2$-dimensional AdS spacetime 
\bena
ds^2 = \frac{\ell^2}{\sin^2 x}\left(-dt^2 + dx^2 \right) \,, 
\label{metric:ads2} 
\eena
which is just the space of orbits $(O^2,g_{ab})$ 
of $(n+2)$-dimensional AdS spacetime 
under the rotation group $SO(n+1)$. 

In the next section, we will perform the spherical harmonic 
decomposition of the equations of motion for scalar fields,
electromagnetic fields and gravitational perturbations and show that
they can be written in the form of a Klein-Gordon equation in
$2$-dimensional AdS spacetime. The analysis of self-adjoint extensions
of the operator $A$ appearing in this wave equation and the relation
between self-adjoint extensions and boundary conditions at infinity
are given in Section~\ref{sect:3}.

\subsection*{Notation and conventions}

We will generally follow the notation and conventions of
\cite{Wald1984}. However, in this paper we will make frequent use of
the above noted fact that $(n+2)$-dimensional AdS spacetime, 
$(M, g_{\mu \nu})$, can be
expressed as a warped product of an $n$-dimensional round unit sphere,
$(S^n, \gamma_{ij})$, and a $2$-dimensional AdS spacetime, $(O^2,
g_{ab})$. To distinguish between
tensors on these different spaces, we will use greek indices to denote
tensors on $M$, we will use latin
indices in the range $a,b,\dots,h$ to denote tensors on $O^2$,
and we will use latin indices in the range $i,j,\dots,p$ to denote
tensors on $S^n$. When convenient, we will also use
these indices to denote coordinates and coordinate components in these
spaces. Thus, we may occasionally write equations like 
eq.~(\ref{metric:ads}) in the form
\bena
 g_{\mu \nu}dx^\mu dx^\nu = g_{ab}(y) dy^ady^b 
+ r^2(y)\gamma(z)_{ij}dz^idz^j \,, 
\label{metric:ads3} 
\eena
where $r \equiv \ell \cos x/\sin x$. 

We will denote the metric compatible derivative operators on $(M,
g_{\mu \nu})$, $(S^n, \gamma_{ij})$, and $(O^2, g_{ab})$ by
$\nabla_\mu$, $\BD_i$, and $\OD_a$, respectively. In order to 
avoid introducing additional notation, we will also use $\gamma_{ij}$
and $\BD_i$ to denote the metric and derivative operator of the general
Riemannian manifolds considered in the next section. In a few
instances, we will need to introduce the derivative operator
associated with the hypersurfaces, $\Sigma$, orthogonal to the static
Killing field. In order to avoid introducing additional notation, we will
also use $D_i$ to denote this derivative operator (as we already did
in eq.~(\ref{ope:differential}) above).

\section{Field equations in anti-de Sitter spacetime} 

In this section, we will analyse the equations of motion for scalar,
electromagnetic, and gravitational perturbations of AdS spacetime. Our
analysis will be based on the decomposition of the modes of these
fields into their scalar, vector, and tensor parts as well as the
expansion of these parts in appropriate spherical harmonics. The
desired general decomposition theorems are derived in the first
subsection, and the spherical harmonic expansion is discussed in the
second subsection. The explicit form of the scalar, electromagnetic,
and gravitational perturbation equations in AdS spacetime are then
derived in the final subsection. Most of the results of
subsections 2.1 and 2.2 are well known and are given in many
references (see, e.g., \cite{Bardeen1980,KS1984,KI2003c}), but
complete derivations/proofs have not usually been provided 
in these other references.

\subsection{Decomposition of Vectors and Symmetric Tensors 
on Compact Manifolds}

We begin by recalling some general results on elliptic operators on
compact manifolds. Let $({\cal M}, \gamma_{ij})$ be a compact Riemannian
manifold. Let $P$ be a differential operator of order $m$ (with smooth
coefficients) mapping tensor fields of type $(k,l)$ on ${\cal M}$ to tensor
fields of the same type on ${\cal M}$. Thus, $P$ has the explicit form
\bena 
P = P^{i_1 \dots i_k}{}_{j_1 \dots j_l \; m_1 \dots m_k}{}^{n_1 
\dots n_l\;p_1 \dots p_m}\BD_{p_1} \dots \BD_{p_m} + 
{\rm lower \,\, order \,\, terms}
\label{P} 
\eena
where $\BD_i$ denotes the derivative operator associated with $\gamma_{ij}$.
For each $x \in {\cal M}$ and each covector $k_i$ at $x$, the symbol, 
$\sigma_P(x,k)$, of $P$ is defined as the following
map from tensors of type $(k,l)$
at $x$ to tensors of type $(k,l)$ at $x$
\bena 
\sigma_P(x,k) = P^{i_1 \dots i_k}{}_{j_1 \dots j_l \; m_1 \dots m_k}{}^{n_1 
\dots n_l\;p_1 \dots p_m}k_{p_1} \dots k_{p_m} \,.
\label{sP} 
\eena 
The operator $P$ is said to be {\it elliptic} if the map
$\sigma_P(x,k)$ is invertible for all $x \in {\cal M}$ and all 
$k_i \neq 0$. A well known result (usually referred to as ``elliptic 
regularity'') states the following: Let $P$ be elliptic and let $j$ be 
a smooth tensor field of type $(k,l)$ (where here and in the following
we will generally suppress all indices). Let $u$ be a distributional tensor
field of type $(k,l)$ which satisfies the equation
\bena 
Pu = j
\label{Pu}
\eena
in the distributional sense. Then $u$ is smooth.

We can use the metric, $\gamma_{ij}$ to define a natural $L^2$
inner product on (complex) tensor fields of type $(k,l)$ on ${\cal M}$ via
\bena 
(u,w)_{L^2} = \int_{\cal M} u^{* i_1 \dots i_k}{}_{j_1 \dots j_l} 
{w_{i_1 \dots i_k}}^{j_1 \dots j_l} \,, 
\eena
where the star denotes complex conjugation. (In this equation
indices are raised and lowered with $\gamma_{ij}$ and the 
integration is with respect to the volume element associated with 
$\gamma_{ij}$.) We may view any differential operator, $P$, as an operator 
on $L^2$, defined on the dense domain of smooth tensor fields. Now 
let $P$ be an elliptic operator on ${\cal M}$ that is symmetric when viewed
as an operator on $L^2$, i.e., for all smooth $u$ and $w$
\bena 
(u, Pw) = (Pu, w) \,. 
\eena
As will be discussed further in subsection 3.1 below,
the self-adjoint extensions of $P$ are determined by the 
``deficiency subspaces'', i.e., the vectors $z_{\pm}$ in 
$L^2$ that satisfy
\bena 
P^* z_{\pm} = \pm i z_{\pm} \,, 
\label{Padjoint}
\eena
where $P^*$ denotes the adjoint of $P$. However, eq.~(\ref{Padjoint})
implies that $z_{\pm}$ satisfies the equation $(P \mp i) z_{\pm}=0$ 
in the distributional sense,
so by elliptic regularity $z_{\pm}$ must be smooth. Hence,
$z_{\pm}$ lies in the domain of $P$, and we may replace $P^*$ by 
$P$ in eq.~(\ref{Padjoint}). Taking the inner product of this equation with 
$z_{\pm}$, we immediately obtain a contradiction unless $z_{\pm}=0$. 
Consequently, the deficiency subspaces are always trivial, and any 
symmetric, elliptic operator on a compact manifold is always essentially 
self-adjoint, i.e., it has a unique self-adjoint extension. Thus, the usually
difficult analysis of determining whether a symmetric 
operator admits self-adjoint extensions and of characterising the self-adjoint
extensions that do exist (see section 3 below) is always trivial for
elliptic operators on a compact manifold.

Now let $P$ be any symmetric, elliptic operator
on ${\cal M}$ that is bounded from below as an operator on $L^2$. By 
the previous remarks, $P$ has a unique self-adjoint
extension, which we shall also denote by $P$. The operator $\exp(-sP)$
can be then defined by the spectral theorem, and, for $s>0$, it is a
bounded self-adjoint operator. Since $P\exp(-sP)$ also is bounded
operator on $L^2$, it follows that $\exp(-sP)$ also is a bounded map
from $L^2$ into the $m$th Sobolev space, $H_m$ of ${\cal M}$, where $m$ is
the order of $P$. Since the embedding of $H_m$ into $L^2$ is a compact
map when ${\cal M}$ is compact 
(see, e.g., \cite{gilkey}), it follows that $\exp(-sP)$ is
a compact, self-adjoint operator on $L^2$. Hence, by the
Hilbert-Schmidt theorem \cite{RS1975}, it admits an orthonormal basis of
eigenvectors, with eigenvalues converging to zero. It follows
immediately that $P$ itself has a purely discrete spectrum, and that
all of the eigensubspaces of $P$ are finite dimensional. Note also that by
elliptic regularity, all eigenvectors of $P$ are smooth tensor
fields. Consequently, an equation of the form (\ref{Pu}) will admit a
solution if and only if $j$ is orthogonal to ${\rm ker} P$, where
${\rm ker} P$ is the subspace of solutions to $Pz =
0$. Furthermore, if a solution to (\ref{Pu}) 
exists, it obviously is unique up to
addition of vectors in ${\rm ker} P$. As already stated above, we have
${\rm dim}({\rm ker} P) < \infty$.

We now are ready to state our two decomposition results. The first one is
essentially a simple case of the Hodge decomposition theorem for differential 
forms.

\medskip  
\noindent
{\bf Proposition 2.1}: Let $({\cal M}, \gamma_{ij})$ be a compact Riemannian
manifold. Then any dual vector field, $v_i$ on ${\cal M}$ can be uniquely
decomposed as
\bena
   v_i =  V_i + \BD_iS   
\label{dec:vi} 
\eena
where $\BD^i V_i = 0$. We refer to $V_i$ and $S$, respectively, as 
the vector and scalar components of $v_i$.

\noindent
{\bf Proof}: If the decomposition exists, then $S$ would have to satisfy
\bena
  \BD^i \BD_i S = \BD^i v_i \,.
\eena
Since $P = -\BD^i \BD_i$ is a positive elliptic operator, we can solve this 
equation for $S$ if and only if the ``source term'' on the
right side is orthogonal 
to ${\rm ker} P$. However, ${\rm ker} P$ consists of only constant
functions and since $\int_{\cal M} \BD^i v_i = 0$, the source is indeed 
orthogonal to ${\rm ker} P$. Hence, we obtain a solution $S$, 
which is unique up to the addition of a constant. We obtain the desired 
decomposition by defining $V_i = v_i - \BD_i S$. This decomposition is 
clearly unique,
since the only ambiguity in $S$ is the addition of a constant, which does not
affect the decomposition. \hfill $\Box$

\medskip

The second proposition concerns the decomposition of symmetric, second
rank tensor fields.

\medskip  
\noindent
{\bf Proposition 2.2}: Let $({\cal M}, \gamma_{ij})$ be a compact Riemannian
Einstein space, i.e., $R_{ij} = c \gamma_{ij}$ for some constant $c$. 
Then any second rank symmetric tensor field, $t_{ij}$ on ${\cal M}$ 
can be uniquely decomposed as
\bena
    t_{ij} &=& T_{ij} 
\non \\
           && + \BD_{(i} V_{j)} 
\non \\ 
           && + \left(
                  \BD_i \BD_j 
                  - \frac{1}{n}\gamma_{ij} \BD^m \BD_m 
                \right) S 
              + \f(1/n)\gamma_{ij} t^m{}_m \,,    
\label{tdec}
\eena 
where $\BD^i T_{ij} = 0$, $T^i{}_i = 0$, and $\BD^i V_i = 0$.
We refer to $T_{ij}$, $V_i$, and $(S, t^m{}_m)$, respectively, 
as the tensor-, vector- and scalar-type components of $t_{ij}$. 

\noindent
{\bf Proof}: We define
\bena 
 J_{ij} &\equiv& t_{ij} - \f(1/n)\gamma_{ij}t^m{}_m \,,   
\\ 
  J_j &\equiv& \BD^i J_{ij} \,,
\label{eq:DiWij=Jj}
\\ 
 J   &\equiv&  \BD^i J_i = \BD^i \BD^j J_{ij} \,.  
\label{eq:DiDjWij=J}
\eena
Consider the equation
\bena
  \left( \BD^i \BD_i + \frac{n}{n-1} c \right) W = \frac{n}{n-1} J \,. 
\label{W}
\eena
This equation can be solved for $W$ if and only if the source $J$ 
is orthogonal to the solutions to
\bena
  \left( \BD^i \BD_i + \frac{n}{n-1} c \right) U = 0 \,. 
\label{U}
\eena
Taking the derivative of this equation and using $R_{ij} = c \gamma_{ij}$, 
we find
\bena
  \left( \BD^i \BD_i + \frac{1}{n-1} c \right) \BD_j U = 0 \,. 
\eena
Contracting with $\BD^j U$ and integrating over ${\cal M}$, 
we find that any solution, 
$U$, to eq.~(\ref{U}) must also satisfy
\bena
  \BD_i \BD_j U - \frac{1}{n} \gamma_{ij} \BD^m \BD_m U = 0 \,. 
\label{U2}
\eena
i.e., $\BD^i U$ must be a conformal Killing field. It is easily checked that
any solution, $U$, to eq.~(\ref{U2}) is automatically orthogonal to $J$, so
eq.~(\ref{W}) always can be solved for $W$. If $c \neq 0$, integration of 
eq.~(\ref{W}) yields $\int_{\cal M} W = 0$. If $c = 0$, then 
$U = {\rm const}$ solves (\ref{U}), 
so we can always choose the solution $W$ to eq.~(\ref{W}) to 
satisfy $\int_{\cal M} W = 0$. Therefore, in all cases, 
we can solve the equation 
\bena
\BD^i \BD_i S = W 
\eena
to obtain $S$. 

Next, we consider the equation
\bena
 \BD^i \BD_{(i} V_{j)} = J_j -\frac{n-1}{n} \BD_j W - c \BD_j S \,. 
\label{Vj}
\eena
Again, the left side is a symmetric elliptic operator on dual vector fields,
so we can solve eq.~(\ref{Vj}) provided that the right side is orthogonal to
the kernel of this operator. However, it is easily seen that the 
kernel consists precisely of Killing vector fields and that the right side
is orthogonal to any Killing vector field. Thus, we can always solve 
eq.~(\ref{Vj}) to obtain $V_j$. Furthermore, taking the divergence of 
eq.~(\ref{Vj}), we find
\bena
\BD^j \BD^i \BD_{(i} V_{j)} = J -\frac{n-1}{n} \BD^j \BD_j W 
                                - c \BD^j \BD_j S = 0 
\label{DVj}
\eena
and hence
\bena
 \left( \BD^i \BD_i + c \right) \BD^j V_j = 0 \,.
\label{DVj2}
\eena
Taking the gradient of this equation, we find
\bena
 (\BD^i \BD_i) \BD_m (\BD^j V_j) = 0 
\label{DVj3}
\eena
which implies that $\BD_i \BD_m \BD^j V_j = 0$. In particular, we have 
$\BD^i \BD_i (\BD^j V_j) = 0$, which implies that $\BD^j V_j$ is constant. 
However, since $\int_{\cal M} \BD^j V_j = 0$, 
we obtain the desired result that $\BD^j V_j = 0$. 

Having now defined $S$ and $V_j$, we simply define $T_{ij}$ by 
eq.~(\ref{tdec}). It is then easily verified that $T^i{}_i = 0$
and that $\BD^i T_{ij} = 0$. Finally, uniqueness of the decomposition 
follows from the fact that each of the 4 terms in eq.~(\ref{tdec}) are 
manifestly orthogonal in the $L^2$ inner product. Consequently, if there were
two distinct ways of writing the decomposition (\ref{tdec}) of $t_{ij}$,
their difference would yield a formula for the zero-vector, expressing it
as the sum of 4 orthogonal vectors, not all of which are 
zero. \hfill $\Box$ 

Finally, we note that compactness of ${\cal M}$ was used in the arguments 
of this subsection in the following two essential ways: (1) to argue
that any symmetric elliptic operator is essentially self-adjoint, and 
(2) to argue that the spectrum of any self-adjoint elliptic operator is 
discrete. The symmetric, elliptic operators considered in the above
propositions will be essentially self-adjoint on many non-compact manifolds
of interest, but their spectra typically will be continuous. 
Consequently, although analogous decomposition results should hold for many 
non-compact manifolds, it will not be possible to give such a 
simple, general proof, i.e., the decomposition results for 
non-compact manifolds must be analysed on a case-by-case basis.

\subsection{Spherical Harmonic Decomposition on an $n$-Sphere}

Proposition 2.1 of the previous subsection holds for arbitrary compact
Riemannian manifolds, and Proposition 2.2 holds for arbitrary compact
Riemannian Einstein spaces. However, on general spaces, an operator 
of interest may ``mix'' the various components of vectors 
and tensors, e.g., when acting on the purely ``tensor part'' 
of a symmetric rank-two tensor 
field, the resulting tensor field could have scalar and vector
components. However, we shall see below
that such ``mixing'' cannot occur on the (round)
$n$-sphere for any operator that is rotationally invariant (and, in
the case of the $2$-sphere, also parity invariant). Consequently, the
decompositions of the previous subsection are particularly useful in
this case, since one can then analyse the scalar, vector, and tensor
components separately. The purpose of this subsection is to establish
this result as well as to introduce the spherical harmonic expansion
of each component.

Let $\gamma_{ij}$ be the round metric on the $n$-sphere,
$S^n$. Then $(S^n, \gamma_{ij})$ is an Einstein space with $c = n-1$, so,
in particular, Proposition 2.2 applies.
Consider, first, the scalar spherical harmonics. The rotation
group $SO(n+1)$ acts naturally on the Hilbert space $L^2_S$ of square
integrable functions $f:S^n \rightarrow {\Bbb C}$. Since the Laplacian,
$\BD^i \BD_i$, is invariant under rotations, each eigensubspace of
$\BD^i \BD_i$ must also be invariant under rotations.
The scalar spherical harmonics, ${\Bbb S}_{{\bf k}_S}$ are defined to be
an orthonormal basis of eigenvectors of $\BD^i \BD_i$, i.e.,
\bena
  \left( \BD^i \BD_i + k_S^2 \right) {\Bbb S}_{{\bf k}_S} =0 \,,  
\label{LS}
\eena 
and 
\bena 
 \int_{S^n} 
                   \scalar_{{\bf k}_S} \scalar_{{\bf k'}_S} 
  &=&   \delta_{{\bf k}_S,{\bf k'}_S}\,.
\eena 
An explicit choice of ${\Bbb S}_{{\bf k}_S}$ can be found in 
e.g., \cite{Higuchi1987} and references therein. 
The eigenvalues $k_S^2$ appearing in eq.~(\ref{LS}) are
$k_S^2 = l(l+n-1),\,\, l=0,1,2,\dots $. 

The rotation group also acts naturally on the Hilbert space of square 
integrable vector fields on $S^n$. In addition, it is easily seen that
rotations map the Hilbert subspace, $L^2_V$, of divergence-free vector fields
into itself. Furthermore, it is easily checked that the Laplacian,
$\BD^i \BD_i$, also maps $L^2_V$ into itself. 
Vector spherical harmonics, ${\Bbb V}_{{\bf k}_V i}$, are defined to be
an orthonormal basis of eigenvectors of $\BD^i \BD_i$ in $L^2_V$, i.e., 
they satisfy
\bena
  (\BD^j \BD_j + k_V^2) {\Bbb V}_{{\bf k}_Vi} =0 \,, \quad 
  \BD^i {\Bbb V}_{{\bf k}_Vi} = 0 \,,   
\eena 
and
\bena
 \int_{S^n} \vector_{{{\bf k}}_Vi} \vector_{{\bf k'}_V}{}^i 
  &=& \delta_{{\bf k}_V,{\bf k'}_V}\,.
\label{LV}   
\eena
An explicit choice of ${\Bbb V}_{{\bf k}_V i}$ can be found in 
e.g., \cite{Higuchi1987} and references therein. 
The eigenvalues $k_V^2$ appearing in eq.~(\ref{LV}) are
$k_V^2 = l(l+n-1)-1,\,\, l=1,2,\dots$. 
The number of the independent components of $\vector_{{\bf k}_V i}$ 
is $(n-1)$, so vector spherical harmonics are nontrivial 
only when $n \geqslant 2$. 

Finally, the rotation group also acts naturally on the Hilbert space
of square integrable rank-two symmetric tensor fields on $S^n$. In
addition, it is easily seen that it maps the Hilbert subspace,
$L^2_T$, of trace-free, divergence-free symmetric tensor fields 
into itself. Furthermore, it
is easily checked that the Laplacian, $\BD^i \BD_i$, also maps $L^2_T$
into itself. Tensor spherical harmonics, $\tensor_{{\bf k}_T ij}$ 
are defined to be an orthonormal basis of 
eigenvectors of $\BD^i \BD_i$ in $L^2_T$, 
i.e., they satisfy
\bena
  (\BD^m \BD_m + k_T^2) \tensor_{{\bf k}_T ij} =0 \,, \quad 
  \BD_j \tensor_{{\bf k}_T}{}^j{}_{i}= 0 \,, \quad 
  \tensor_{{\bf k}_T}{}^i{}_{i} = 0 \,,    
\label{LT}
\eena 
and 
\bena 
 \int_{S^n} 
   \tensor_{{\bf k}_T ij} \tensor_{{\bf k'}_T}{}^{ij}  
  &=& \delta_{{\bf k}_T,{\bf k'}_T} \,.   
\eena 
An explicit choice of $\tensor_{{\bf k}_T ij}$ can be found  
in Ref.~\cite{Higuchi1987}.
The eigenvalues $k_T^2$ appearing in eq.~(\ref{LT}) are
$k_T^2 = l(l+n-1)-2,\,\, l=2,3,\dots$. 
The number of the independent components of $\tensor_{{\bf k}_T ij}$ 
is $ (n-2)(n+1)/2$, so, in particular,
tensor spherical harmonics are nontrivial only when $n \geqslant 3$.

The ``total squared angular momentum'', $J^2$, is the Casimir element of the 
universal enveloping algebra of the Lie algebra of $SO(n+1)$ obtained
by summing the squares of the elements of an orthonormal basis, 
$\{\xi_{(\alpha)}\}$, of the Lie
algebra. For tensor fields on $S^n$, the action of $J^2$ is obtained by
replacing each $\xi_{(\alpha)}$ by the Lie derivative with respect to
the corresponding Killing vector field on $S^n$. The action of $J^2$ on
scalar functions on $S^n$ is thereby computed to be simply
\bena
J^2_S = -\BD^i \BD_i \,. 
\label{JS}
\eena
Similarly, the action of $J^2$ on dual vector fields and on rank-two
symmetric tensor fields on $S^n$ are, respectively,
\bena
(J^2_V)_i{}^j = - (\BD^m \BD_m - (n-1)) \delta_i{}^j \,, 
\label{JV}
\eena
and
\bena
(J^2_T)_{ij}{}^{kl} = - (\BD^m \BD_m - 2n) \delta_i{}^k \delta_j{}^l
-2 \gamma_{ij} \gamma^{kl} \,. 
\label{JT}
\eena 
Consequently, for scalar spherical harmonics of order $l$, $J^2$ takes
the value $l(l+n-1)$; for vector spherical harmonics of order $l$,
$J^2$ takes the value $l(l+n-1) + (n-2)$; and for tensor spherical
harmonics of order $l$, $J^2$ takes the value $l(l+n-1) +2(n-1)$. By 
inspection, the values of $J^2$ for scalar, vector, and tensor harmonics
never are equal, except for the case $n=2$, when the values of $J^2$ for 
scalar and vector spherical harmonics agree for all $l$. Thus---except 
in the case $n=2$---the scalar, vector, and tensor harmonics
always span inequivalent representations of the rotation group. 
This implies that there cannot exist any (nonzero) 
rotationally invariant map between any pair of the spaces
$L^2_T$ and $L^2_V$ and $L^2_S$---with the exception 
of the case $n=2$, where there could exist a rotationally invariant map 
between $L^2_V$ and $L^2_S$ (see below). In particular, this means that any 
rotationally invariant operator on rank-two
symmetric tensor fields on $S^n$ cannot take the tensor
component into a nonzero vector or scalar component; it cannot take the 
vector component into a nonzero tensor component or (if $n > 2$)
a nonzero scalar component; nor can it take the 
scalar component into a nonzero tensor component or (if $n > 2$) 
a non-zero vector component.

When $n=2$ the scalar and vector harmonics span equivalent representations
of $SO(3)$ and, indeed, the vector harmonics can be chosen to be
\bena
   \vector_{{\bf k}i} \equiv \epsilon_{ij} \BD^j\scalar_{\bf k} \,,  
\label{SV}
\eena 
where $\epsilon_{ij}$ is the (positively oriented) 
volume element on $S^2$ associated 
with $\gamma_{ij}$. Thus, one sees explicitly that
there exist nonvanishing $SO(3)$-invariant
maps between $L^2_V$ and $L^2_S$. However, as can be seen from
eq.~(\ref{SV}), the vector and scalar harmonics behave differently under
parity, and thus span inequivalent representations of $O(3)$. 
(The scalar representations of $O(3)$ are said to be of ``polar'' (or 
``even parity'') type, and the vector representations are said to
be of ``axial'' (or ``odd parity'') type.)
Consequently,
even when $n=2$, no rotationally {\it and} parity invariant operator can
``mix'' the scalar and vector components of a vector field or symmetric,
rank-two tensor field.

\subsection{Scalar fields on AdS Spacetime}  

We now turn our attention to the derivation of the desired form of the 
field equations for scalar, electromagnetic, and gravitational perturbations
of AdS spacetime. As previously noted, the metric (\ref{metric:ads}) 
of $(n+2)$-dimensional AdS spacetime
takes the form of a warped product of an $n$-dimensional
base space $(S^n,\gamma_{ij})$ and a
$2$-dimensional space of orbits $(O^2,g_{ab})$: 
\bena
 ds^2_{(n+2)}= g_{\mu \nu}dx^\mu dx^\nu 
             = g_{ab}(y) dy^ady^b+r^2(y)\gamma(z)_{ij}dz^idz^j\,,    
\label{metric:warped-product} 
\eena  
where $\gamma_{ij}$ is the round, unit sphere metric on $S^n$, and
\bena
r \equiv \ell \frac{\cos x}{\sin x}\,.
\eena
In the coordinates $(t,r)$ 
the metric of the space of orbits takes the form 
\bena
 g_{ab}(y)dy^ady^b = - V^2 \ell^2 dt^2 + \frac{dr^2}{V^2} \,, \quad 
 V^2 = 1+\frac{r^2}{\ell^2} \,. 
\eena
Our general strategy will be to project all tensor fields into parts 
that are tangent and orthogonal to $S^n$, to decompose all tensor fields
on $S^n$ into their scalar, vector, and tensor components, and to expand
each component on $S^n$ in the appropriate spherical harmonics. 

To avoid confusion when comparing with other analyses, it should be
emphasised that our analysis involves an ``$n+2$'' decomposition of
AdS spacetime rather than an ``$(n+1) + 1$'' decomposition. Thus, for
example, our notion of the scalar, vector, and tensor components of a
metric perturbation are defined with respect to the action of the
isometry subgroup $SO(n+1)$, whose orbits are $n$-spheres. It would
also be possible to write AdS spacetime as a Robertson-Walker
spacetime and decompose tensors with respect to the isometry subgroup
acting on the $(n+1)$-dimensional surfaces of homogeneity and
isotropy in the manner done, e.g., in \cite{Bardeen1980,KS1984}. 
This would give rise to completely different notions of the scalar, 
vector, and tensor components of a metric perturbation. 

The simplest case is that of a Klein-Gordon scalar field, $\phi$, 
where no projections are necessary. We expand $\phi$ in terms of 
scalar spherical harmonics $\scalar_{\bf k}(z^i)$ 
\bena 
\phi(t,x,z^i)
\equiv r^{-n/2} \sum_{{\bf k}_S}\Phi_{\bf k}(t,r) 
                                \scalar_{\bf k}(z^i) \,.
\label{redef} 
\eena 
The equation of motion for each $\Phi_{\bf k}(t,r)$ is obtained by 
straightforward substitution into (\ref{eq:Klein-Gordon}). Returning
to the original coordinates $(t,x)$, we obtain
\bena 
 { \partial^2 \over \partial t^2}\Phi  
 = \left(
      \frac{ \partial^2}{ \partial x^2}
     - \frac{\nu^2-1/4}{\sin^2x} 
     -\frac{\sigma^2-1/4}{\cos^2x}  
   \right) \Phi \,,  
\label{eq:canonical} 
\eena   
where the index ${\bf k}$ labelling 
the harmonics has been omitted. Here the constants 
$\nu$ and $\sigma$ in the effective      
potential term are given by 
\bena 
   \nu^2 -\frac{1}{4} = \frac{n(n+2)}{4} + m_0^2 \ell^2 \,, 
\label{def:nu}
\eena 
and 
\bena
  \sigma^2 - \frac{1}{4} = l(l+n-1) + \frac{n(n-2)}{4} \,.
\label{def:sigma}  
\eena  

Since the scalar curvature, $R = - (n+1)(n+2)\ell^{-2}$, is constant
in AdS spacetime, the modification to eq.~(\ref{eq:Klein-Gordon}) 
resulting from  
the addition of the curvature coupling term $-\xi R$ is equivalent to changing
the mass by $m_0^2 \rightarrow m^2$, where 
the effective mass, $m^2$, is defined by
\bena
m^2 \equiv m^2_0 - \xi (n+1)(n+2)\ell^{-2} \,. 
\label{meff}
\eena
Note that
for the case of a conformally coupled scalar field (i.e., $m_0^2 = 0$ and
$\xi = {n}/{4(n+1)}$), we have  
\bena
  \nu^2= \frac{1}{4} \,.  
\label{mu:conformally-invariant}   
\eena 

\subsection{Electromagnetic fields}   
Maxwell's equations for the electromagnetic
field tensor $F_{\mu \nu}$ are 
\bena
&& \nabla_\nu F^{\nu\mu} =0 \,, 
\label{eq:eom} 
\\
&&
\nabla_{[\sigma} F_{\mu \nu]} =0 \,. 
\label{eq:Bianchi}
\eena 
Since AdS spacetime is topologically trivial
the second equation~(\ref{eq:Bianchi}) immediately 
implies the existence of a vector potential $A_\mu$, so that 
\bena 
 F_{\mu \nu} = \nabla_\mu A_\nu - \nabla_\nu A_\mu \,. 
\label{F-A} 
\eena 
Equation (\ref{F-A}) is the general solution to eq.~(\ref{eq:Bianchi}).


We may write $A_\mu$ as the sum of its projection orthogonal to $S^n$ and its 
projection tangent
to $S^n$. The components of $A_\mu$ orthogonal to $S^n$ 
behave as scalars under rotations of $S^n$. The projection 
of $A_\mu$ tangent to 
$S^n$ can then be decomposed into its 
scalar and vector parts in accordance with
Proposition 2.1. Maxwell's equations are invariant under rotations and parity
transformations, and hence cannot couple the vector and scalar parts of 
$A_\mu$. Therefore, we may treat these cases separately.

In order to express Maxwell equation (\ref{eq:eom}) explicitly, 
it is useful to write down the non-vanishing components 
of the Christoffel symbol $\Gamma^\mu{}_{\nu \lambda}$ 
associated with $g_{\mu \nu}$, 
\bena
  \Gamma^a{}_{bc} = \widehat{\Gamma}^a{}_{bc} \,, \quad 
  \Gamma^a{}_{ij} = - r\OD^ar \gamma_{ij} \,, \quad 
  \Gamma^i{}_{aj} = \frac{\OD_ar}{r}\delta^i{}_j \,, \quad 
  \Gamma^i{}_{jk} = \hat{\Gamma}^i{}_{jk} \,, 
\eena 
where $\widehat{\Gamma}^a{}_{bc}$ and $\hat{\Gamma}^i{}_{jk}$ are  
the components of the Christoffel symbols associated with the metrics 
$g_{ab}$ and $\gamma_{ij}$, respectively. 

\subsubsection*{Vector-type component}  

The purely vector part, $A^V_\mu$, of $A_\mu$ can be 
expanded in vector spherical harmonics as 
\bena 
 A^V_\mu dx^\mu 
 = \sum_{{\bf k}_V} \phi_{V{\bf k}}\vector_{{\bf k}i} dz^i\,.
\label{AV}
\eena
Since the gauge freedom in $A_\mu$ is purely scalar, 
it follows that $A^V_\mu$---and 
hence each $\phi_{V{\bf k}}$---is gauge-invariant. Substituting 
eq.~(\ref{AV}) into eq.~(\ref{eq:eom}), we obtain 
\bena 
 \f(1/r^{n-2})\OD_c (r^{n-2} \OD^c \phi_{V}) 
  - \frac{k_V^2 + (n-1)}{r^2} \phi_{V} = 0 \,,  
\eena 
where we have omitted the index ${\bf k}$ here and in the following.
In terms of the rescaled variable $\Phi_V \equiv r^{(n-2)/2}\phi_V$, 
we obtain 
\bena
  \OD^a\OD_a \Phi_V - \frac{n-2}{2}\left\{ 
                                   \frac{\OD^a\OD_a r}{r} 
                                 + \frac{n-4}{2} \frac{(\OD^cr)\OD_cr}{r^2} 
                             \right\} \Phi_V
                          -  \frac{k_V^2 + (n-1)}{r^2} \Phi_V = 0\,. 
\label{eq:Max-vect-3}
\eena 
We can easily check that the $a$-components of eq.~(\ref{eq:eom}) 
provide just a trivial equation, as they must since they are scalar
components of Maxwell's equations. 

Substituting $\OD^a\OD_a r/r = 2/\ell^2$ and 
${(\OD^cr)\OD_cr}/{r^2} = 1/\ell^2 +1/r^2$ 
into eq.~(\ref{eq:Max-vect-3}) we obtain
\bena 
 \OD^a\OD_a \Phi_V 
    - \left\{ 
              \frac{n(n-2)}{4}\frac{1}{\ell^2}  
             + \left[\frac{n(n-2)}{4}+l(l+n-1)\right]\f(1/r^2)
      \right\} \Phi_V = 0 \,.
\label{maxwel:vector}
\eena
In terms of the original coordinates
$(t,x)$, eq.~(\ref{maxwel:vector}) takes exactly the same form as
eq.~(\ref{eq:canonical}), with
\bena
  \nu^2 &=& \frac{n(n-2)}{4} +\frac{1}{4} \,, 
\label{em:nu:vector}
\eena 
and with $\sigma^2$ again given by eq.~(\ref{def:sigma}). 

\subsubsection*{Scalar-type component}   

The scalar part, $A^S_\mu$, of $A_\mu$, consists of two quantities: 
a vector, $A_a$, orthogonal to $S^n$ and the scalar component, $A$, of a
vector tangent to $S^n$ (see Proposition 2.1). We may expand both of these
quantities in scalar spherical harmonics $\scalar_{\bf k}$, 
\bena 
 A^S_\mu dx^\mu 
 =  \sum_{{\bf k}_S}
                \left(
                   A_{{\bf k}a} \scalar_{\bf k} dy^a 
                  +A_{\bf k} \BD_i \scalar_{{\bf k}} dz^i
                \right) \,, 
\eena
The contribution of each spherical harmonic component to
the field strength is given by 
\bena
&&  F_{ab} = (\OD_aA_b-\OD_bA_a)\scalar \,, 
\label{scalar:Fab}
\\
&&  F_{ai} = (\OD_aA + k_S A_a) \BD_i \scalar \,,
\label{scalar:Fai}
\\
&&  F_{ij} = 0 \,,  
\label{scalar:Fij}  
\eena
where we have again omitted writing the index ${\bf k}$. 
The $i$-components of eq.~(\ref{eq:eom}) then take the form
\bena 
 \OD_a \left\{ r^{n-2} \left( \OD^aA + k_S A^a \right) \right\} = 0 \,. 
\label{FS}
\eena 
This immediately implies that there exists a scalar field $\phi_S$ such that 
\bena 
 \OD_a \phi_S 
 = \epsilon_{ab}r^{n-2} (\OD^bA + k_S A^b) \,,   
\label{def:phi}
\eena 
where $\epsilon_{ab}$ is the metric compatible volume element 
on $(O^2,g_{ab})$.  
Since $\phi_S$ is related to the field strength via eq.~(\ref{scalar:Fai}) 
as   
\bena
 F_{ai} = \frac{1}{r^{n-2}} \epsilon_{ab} \OD^b \phi_S \BD_i \scalar \,,    
\eena 
gauge freedom of $\phi_S$ is restricted to only 
the replacement 
\bena
  \phi_S \rightarrow \phi_S + {\rm const} \,.   
\label{transf:constant}
\eena
In terms of $\phi_S$, the $a$-components of
eq.~(\ref{eq:eom}) are written as  
\bena 
 \OD_b \left[
           r^n \OD_a \left( \f(1/r^{n-2}) \OD^a \phi_S \right) 
           - k_S^2 \phi_S 
     \right] = 0 \,.      
\label{eq:Max-scal-0}
\eena 
So, we have 
\bena
 r^{n-2}\OD_a \left( \f(1/r^{n-2}) \OD^a \phi_S \right) 
 - \f(k_S^2/r^2)\phi_S = \frac{c}{r^2} \,. 
\label{eq:Max-scal-1}
\eena 
Since the integration constant $c$ in the right hand side of this equation 
can always be absorbed in $\phi_S$ 
by using the remaining gauge freedom (\ref{transf:constant}), 
we have the 
following equation of motion for the now gauge-invariant variable $\phi_S$,  
\bena
  r^{n-2}\OD_a \left( \f(1/r^{n-2}) \OD^a \phi_S \right) 
  - \f(k_S^2/r^2)\phi_S = 0 \,. 
\label{eq:Max-scal-2}
\eena  
In terms of the rescaled variable $\Phi_S \equiv r^{-(n-2)/2}\phi_S$, 
we have
\bena  
  \OD^a\OD_a \Phi_S + \frac{n-2}{2}\left\{ 
                                   \frac{\OD^a\OD_a r}{r} 
                                 - \frac{n}{2} \frac{(\OD^cr)\OD_cr}{r^2} 
                             \right\} \Phi_S
                          -  \frac{k_S^2}{r^2} \Phi_S = 0\,. 
\label{eq:Max-scal-3}
\eena 
Again substituting $\OD^a\OD_a r/r = 2/\ell^2$ and 
${(\OD^cr)\OD_cr}/{r^2} = 1/\ell^2 +1/r^2$, we obtain 
\bena
 \OD^a\OD_a\Phi_S 
    - \left\{ 
             \frac{(n-2)(n-4)}{4}\frac{1}{\ell^2}   
             +\left[\frac{n(n-2)}{4}+l(l+n-1)\right]\f(1/r^2)
      \right\} \Phi_S = 0 \,.  
\label{maxwel:scalar}
\eena 
In terms of the original coordinate system $(t,x)$, 
this equation again takes the canonical
form~(\ref{eq:canonical}) with 
\bena
  \nu^2 &=& \frac{(n-2)(n-4)}{4} +\frac{1}{4} \,, 
\label{em:nu:scalar}
\eena 
and $\sigma^2$ again given by eq.~(\ref{def:sigma}).
Notice that in $4$-dimensional AdS spacetime (i.e., $n=2$), 
both vector- and scalar-type components of the electromagnetic 
field satisfy the same equation as a conformally invariant 
scalar field.

\subsection{Gravitational perturbations} 

We shall basically follow the derivation of equations of motion 
for the master variables of gravitational perturbations 
developed in Ref.~\cite{KIS2000}. 
The resultant equations are equivalent to those given 
in~\cite{Mukoh2000} in maximally symmetric backgrounds 
and those in~\cite{KI2003c,KI2003a} in more general backgrounds.  

Gravitational perturbations are described by a symmetric tensor field 
$h_{\mu \nu} = \delta g_{\mu \nu}$  on the background 
AdS spacetime. We may project $h_{\mu \nu}$ relative to $S^n$ as
\bena
 h_{\mu \nu} dx^\mu dx^\nu 
 = h_{ab}dy^ady^b + 2 h_{ai}dy^adz^i + h_{ij}dz^idz^j \,. 
\label{tensor-on-base}
\eena 
The projection, $h_{ab}$, orthogonal to $S^n$ in both indices 
is purely scalar 
with respect to rotations. The orthogonal-tangent projection, 
$h_{ai}$, 
and the tangent-tangent projection $h_{ij}$ can be further 
decomposed into their scalar, vector, and tensor parts as 
\bena 
 h_{ai} &=& {\BD}_{i}h_a + h^{(1)}{}_{ai} \,,  \quad 
\label{decompose:hai}
\\
 h_{ij} &=& h^{(2)}{}_{ij} 
\non \\
        &&\, + 2 {\BD}_{(i}h^{(1)}_T{}_{j)}
\non \\
        && \, + h_L \gamma_{ij} 
              +  \left(
                       {\BD}_i{\BD}_j 
                       -\frac{1}{n} \gamma_{ij} \BD^m\BD_m 
                 \right) h^{(0)}_T \,,  
\label{decompose:hij}
\eena 
with    
\bena  
&& {\BD}^jh^{(2)}{}_{ij}= h^{(2)}{}^i{}_i = 0 \,,   
\\ 
&&
 {\BD}^i h^{(1)}{}_{ai}= 0 \,, \quad 
 {\BD}^{i}h^{(1)}_T{}_{i}= 0 \,.   
\eena
Thus, the tensor part of $h_{\mu \nu}$ is $h^{(2)}{}_{ij}$, 
the vector part of $h_{\mu \nu}$ consists of 
$(h^{(1)}{}_{ai},h^{(1)}_T{}_{j})$, 
and the scalar component of $h_{\mu \nu}$ consists of 
$(h_{ab}, h_L,h^{(0)}_T)$.  
Each tensorial component decouples from the others and can 
be expanded in the appropriate tensor harmonics, 
namely ${\Bbb T}_{{\bf k}_T ij}$, ${\Bbb V}_{{\bf k}_Vi}$, 
and ${\Bbb S}_{{\bf k}_S}$, respectively.

\subsubsection*{Tensor-type perturbation}  
We expand the tensor-type perturbation in spherical harmonics as, 
\bena 
 h^{(2)}{}_{ij} &=& \sum_{{\bf k}_T} H^{(2)}_{T{\bf k}} \cdot 
                                   \tensor_{{\bf k} ij} \,, 
\label{exp:tensor}
\eena
with $H^{(2)}_{T}$ being a function on $(O^2,g_{ab})$.  
Since the gauge freedom 
\bena
h_{\mu \nu} \rightarrow h_{\mu \nu} 
- \nabla_\mu \xi_\nu - \nabla_\nu \xi_\mu \,,
\label{transf:gauge}
\eena
does not contain any tensor part, it is clear that
$H^{(2)}_T$ is gauge-invariant. Writing $\Phi_T \equiv r^{-2 +n/2}H^{(2)}_T$,
we find that the linearised Einstein equations yield 
\bena  
     \OD^a\OD_a \Phi_T
     - \left\{ 
              \frac{n(n+2)}{4}\frac{1}{\ell^2}  
             + \left[\frac{n(n-2)}{4}+l(l+n-1)\right]\f(1/r^2)
      \right\} \Phi_T =0 \,.  
\label{eq:covariant-master-T} 
\eena  
This equation again takes the form~(\ref{eq:canonical})  
with $\nu^2$ given by
\bena
  \nu^2 = \frac{n(n+2)}{4} +\frac{1}{4} \,, 
\label{def:nu:tensor}
\eena  
and $\sigma^2$ again given by eq.~(\ref{def:sigma}).

\subsubsection*{Vector-type perturbation} 

Vector-type components can be expanded in terms of vector spherical
harmonics as
\bena 
 h^{(1)}{}_{ai} &=&  \sum_{{\bf k}_V} H^{(1)}_{{\bf k}}{}_a \cdot 
                                      \vector_{{\bf k}i} \,,  
\label{exp:vector-1}
\\
  h^{(1)}_T{}_j 
  &=& \sum_{{\bf k}_V} H^{(1)}_{T{\bf k}} \cdot 
                       \vector_{{\bf k}i} \,.  
\label{exp:vector-2}
\eena 

Note that, as shown in the Appendix, $l=1$ modes do not correspond to 
dynamical perturbations, so hereafter we assume 
$l\geqslant 2$.

The vector component, $\xi^{(1)} \vector_idz^i$, of the gauge 
transformation eq.~(\ref{transf:gauge}) yields
\bena
 H^{(1)}{}_a &\rightarrow& H^{(1)}{}_a 
                 - r^2 \OD_a\left(\frac{\xi^{(1)}}{r^2}\right) \,, 
\\
 H^{(1)}_{T} &\rightarrow& H^{(1)}_{T} - \xi^{(1)} \,.  
\eena 
It follows immediately that a natural gauge-invariant combination 
is given by 
\bena
  Z_a \equiv H^{(1)}{}_a
             - r^2 \OD_a \left(\frac{H^{(1)}_T}{r^2}\right)\,.  
\eena  
Substituting this into the linearised Einstein equations, we have
\bena
&& \OD_a\left(r^{n-2} Z^a\right) = 0 \,,  
\label{eq:vector-constraint}
\\
&&  \frac{1}{r^{n}}\OD^b 
  \left[
       r^{n+2}\left\{\OD_b\left(\frac{Z_a}{r^2}\right) 
                     - \OD_a\left(\frac{Z_b}{r^2}\right) 
              \right\} 
  \right] - \frac{k_V^2-(n-1)}{r^2} Z_a = 0 \,.   
\label{eq:vector-evolve} 
\eena 
These equations have the same form as the $i$-components and $a$-components,
respectively, of eq.~(\ref{eq:eom}) in the case of the scalar part of 
electromagnetic fields, with $\OD_aA + k_SA_a$
replaced by $Z_a$. In parallel with the introduction of $\phi_S$
in the electromagnetic case (see eq.~(\ref{FS})), we find from
eq.~(\ref{eq:vector-constraint}), that there exists a scalar 
potential $\phi_V$ such that 
\bena 
   Z_a = \frac{1}{r^{n-2}} \epsilon_{ab} \OD^b \phi_V \,. 
\eena   
Substituting this into equation~(\ref{eq:vector-evolve}), 
we obtain
\bena
  \OD^a\left[
           r^{n+2}\OD_b\left( \frac{1}{r^n}\OD^b\phi_V\right) 
           - \{k_V^2-(n-1)\} \phi_V
     \right] = 0 \,. 
\eena 
{}From this equation (after eliminating an integration 
constant using the degree of freedom $\phi_V \to \phi_V + {\rm const.}$ 
as we have done in the electromagnetic field case), we obtain   
\bena
     \OD^a\OD_a \Phi_V
     - \left\{ 
              \frac{n(n-2)}{4}\frac{1}{\ell^2}  
             + \left[\frac{n(n-2)}{4}+l(l+n-1)\right]\f(1/r^2)
      \right\} \Phi_V =0 \,, 
\label{eq:covariant-master-V} 
\eena 
where $\Phi_V \equiv r^{-n/2} \phi_V$. 
This equation again has the form~(\ref{eq:canonical}) 
with $\nu^2$ given by
\bena
  \nu^2 =  \frac{n(n-2)}{4} +\frac{1}{4} \,, 
\label{def:nu:vector}
\eena  
and $\sigma^2$ again given by eq.~(\ref{def:sigma}).
Note that this equation is precisely the same as the vector-type 
of Maxwell's equations~(\ref{maxwel:vector}).

\subsubsection*{Scalar-type perturbation}  

The scalar components of $h_{\mu \nu}$ can be expanded in scalar spherical
harmonics as follows:   
\bena
 h_{ab} &=& \sum_{{\bf k}_S} H^{(0)}_{{\bf k}}{}_{ab} \cdot 
                             \scalar_{\bf k} \,,
\label{exp:scalar-1}
\\
 h_a &=&   \sum_{{\bf k}_S} H^{(0)}_{\bf k}{}_{a} \cdot 
                            \scalar_{\bf k} \,,
\label{exp:scalar-2}
\\
 h_L &=& \sum_{{\bf k}_S} H^{(0)}_{L{\bf k}} \cdot 
                          \scalar_{\bf k} \,,
\label{exp:scalar-3}
\\ 
 h^{(0)}_T &=&  \sum_{{\bf k}_S} H^{(0)}_{T {\bf k}} \cdot 
                                 \scalar_{\bf k} \,, 
\label{exp:scalar-4}
\eena
where each of the expansion coefficients, 
$(H^{(0)}{}_{ab}, H^{(0)}{}_a, H^{(0)}_L,H^{(0)}_T)$ 
are functions on $(O^2,g_{ab})$.  

Note that the $l=0$ ($k_S^2=0$) mode preserves the spherical symmetry 
of the background and, by Birkhoff's theorem, cannot describe any 
dynamical degrees of freedom. For $l=1$ ($k_S^2=n$), 
the corresponding perturbations also do not describe any physical dynamical 
degrees of freedom, as we show in the Appendix. Hence, hereafter we consider 
only the $l\geqslant 2$ modes. 

The scalar-type gauge transformation is generated by 
$\xi^{(0)}{}_a \scalar dy^a + \xi^{(0)} {\BD}_i \scalar dz^i$. 
Observing that the scalar-type components transform as 
\bena
 H^{(0)}{}_{ab} &\rightarrow& H^{(0)}{}_{ab}
                            -\OD_a\xi^{(0)}{}_b-\OD_b\xi^{(0)}{}_a \,,
\\
 H^{(0)}{}_{a}&\rightarrow& H^{(0)}{}_{a}
                   - \xi^{(0)}{}_a
                   -r^2\OD_a\left( \frac{\xi^{(0)}}{r^2}\right)  \,, 
\\
 H^{(0)}_{L} 
 &\rightarrow&
              H^{(0)}_{L} + \frac{2k_S^2 }{n}\xi^{(0)} 
              - 2r(\OD^ar) \xi^{(0)}{}_a \,,
\\
 H^{(0)}_{T} &\rightarrow& H^{(0)}_{T} - 2  \xi^{(0)} \,, 
\eena 
one can check that the following combinations  
are gauge-invariant: 
\bena  
&& Z \equiv n r^{n-4}
            \left\{             
                   H^{(0)}_L 
                  +\frac{k_S^2}{n}H^{(0)}_T 
                 + 2r(\OD^ar) X_a 
            \right\} \,, 
\label{def:Z}
\\
&& Z_{ab} \equiv 
          r^{n-2} \left(
                        H^{(0)}{}_{ab} + \OD_aX_b + \OD_bX_a 
                  \right)
          + \frac{(n-1)}{n} Z g_{ab} \,, 
\label{def:Zab}
\eena 
where 
\bena 
  X_a \equiv - H^{(0)}{}_a 
             + \frac{1}{2} r^2 \OD_a 
               \left(\frac{H^{(0)}_T}{r^2}\right) \,, 
\eena 
which itself is gauge-dependent: $X_a \to X_a +\xi^{(0)}{}_a$.  

The full set of linearised Einstein equations for the scalar-type 
perturbations are written out in eqs.~(63)---(66) of 
Ref.~\cite{KIS2000}, where one must substitute $F = r^{-n+2} Z/(2n)$ 
and $F_{ab} = r^{-n+2} Z_{ab}- \{(n-1)/n\} r^{-n+2} Zg_{ab}$ 
in order to express these equations in terms of the variables 
we have introduced here.
The $(a,i)$-components and the traceless part of the $(i,j)$-components 
of the linearised Einstein equations yield
\bena
  \OD_b Z^b{}_a &=& \OD_a Z^c{}_c \,,  
\label{einstein:ij}
\\
  Z^a{}_a &=& Z \,.  
\label{einstein:ai} 
\eena 

We will now show that the variable $Z_{ab}$ (and, hence, by 
eq.~(\ref{einstein:ai}), also the variable $Z$) can be replaced 
by a single scalar field, $\phi_S$. To do so, we will want 
$\phi_S$ to satisfy
\bena
  \left( \OD^a\OD_a -\frac{2}{\ell^2} \right)\phi_S =  Z^a{}_a   
\label{Zaa}
\eena 
on the space of orbits $(O^2, g_{ab})$. However, $(O^2, g_{ab})$, 
of course, fails to be globally 
hyperbolic, and although the methods of this paper allow us to 
prove existence of solutions to the homogeneous equation with initial
data in $L^2(\Sigma,V^{-1}d\Sigma)$, we cannot 
guarantee existence of solutions to the inhomogeneous equation with an
arbitrary smooth source that might, in particular, fail to lie in 
$L^2(\Sigma,V^{-1}d\Sigma)$ at each time. Nevertheless, we
can overcome this difficulty by noticing that $(O^2, -g_{ab})$ {\em is}
a globally hyperbolic spacetime, as can be seen by inspection of
eq.~(\ref{metric:ads2}), recalling that the coordinate ranges of $(x,t)$ are 
$-\infty < t < \infty$ and $0 < x \leqslant \pi/2$. Thus, we can solve 
eq.~(\ref{Zaa}) for arbitrary smooth $Z^a{}_a$ by specifying arbitrary
smooth ``initial data'' $(\phi_{S0}(t), (\partial \phi_{S0}/\partial x) (t))$
on a hypersurface $x = x_0$. Furthermore, we will specify this initial data
so as to satisfy
\bena 
 t^b \left( \OD_a\OD_b - \frac{1}{\ell^2} g_{ab}\right)\phi_S = t^b Z_{ab} 
\eena
on the initial data surface $x=x_0$, where $t^a$ denotes 
the static Killing field $(\partial/\partial t)^a$. (The above equation 
yields a system of two linear ODE's in $t$ for 
$\phi_{S0}$ and $\partial \phi_{S0}/\partial x$ and hence always 
can be solved.) Now define
\bena 
 T_{ab} = Z_{ab} 
         - \left( \OD_a\OD_b - \frac{1}{\ell^2} g_{ab}\right)\phi_S
	 \,.    
\eena 
Then $T_{ab}$ is transverse and traceless on $(O^2,g_{ab})$, i.e., 
\bena
 \OD_bT^b{}_a=0 \,, \quad T^a{}_a=0 \,.   
\label{condi:t-t}
\eena 
We define
\bena
   v_a = T_{ab}t^b \,.  
\eena 
It follows immediately
from the Killing's equation $\OD_at_b + \OD_bt_a =0$ and 
eq.~(\ref{condi:t-t}) that
\bena
   \OD_av^a =0 \,, 
\eena 
and, consequently, there exists a scalar field $s$ such that
\bena
    v^a = \epsilon^{ab}\OD_b s \,. 
\label{expr:va:s} 
\eena 
By our above choice of initial data for $\phi_S$, we have $v_a = 0$ at
$x=x_0$, so $\OD_b s = 0$ at $x=x_0$. By adding a constant to $s$ if necessary,
we therefore can ensure that on the ``initial data surface'' $x=x_0$
we have $s = 0$ and $\partial s/\partial x = 0$. 

On the other hand, we also have
\bena
   \OD_a (\epsilon^{ab}v_b)&=& \OD_a (\epsilon^{ab}T_b{}^ct_c)
\non \\
   &=& \OD_a (\epsilon^{cb}T_b{}^at_c)
\non \\
   &=& \epsilon^{cb}T_b{}^a\OD_at_c
\non \\
   &=& -\frac{1}{2}\epsilon^{cb}T_b{}^a
       \epsilon_{ac}\epsilon^{de}\OD_dt_e 
\non \\
   &=&  \frac{1}{2}\delta^b{}_aT_b{}^a
        \epsilon^{de}\OD_dt_e 
\non \\
   &=& 0 \,,  
\label{curlv}
\eena
where we have used the fact that for any traceless tensor $T_{ab}$ 
on $(O^2,g_{ab})$, 
\bena
T^a{}_b\epsilon^{bc} = T^c{}_b\epsilon^{ba}
\label{formula:T-epsilon} 
\eena  
at the second line and the formula 
$\OD_at_c=-(1/2)\epsilon_{ac}\epsilon^{de}\OD_dt_e$ for any Killing vector 
field $t^a$ on $(O^2,g_{ab})$ at the fourth line. 
Substituting eq.~(\ref{expr:va:s}) into eq.~(\ref{curlv}), we obtain
\bena
   \OD^a\OD_a s=0 \,.  
\label{eq:s} 
\eena 
Consequently, by Cauchy evolution in $x$, we obtain $s=0$ and, hence,
$v_a = 0$ throughout $(O^2,g_{ab})$. However, using the formulae 
\bena
  g_{ab} &=& \frac{1}{V^2}(-t_at_b + \epsilon_{ac}\epsilon_{bd}t^ct^d)
  \,,
\label{formula:gab-ttepsepstt}
\\  
 \epsilon_{ab} \epsilon_{cd}t^d &=& t_ag_{bc}- t_bg_{ca} \,,  
\label{formula:eps-eps-t} 
\eena 
we find 
\bena
  T_{ab} &=& T_a{}^cg_{cb} 
\non \\
    &=& 
       \frac{1}{V^2}
       \left( 
             -T_a{}^ct_ct_b 
             +T_d{}^c\epsilon_{ca}\epsilon_{be}t^dt^e \right) 
\non \\
    &=& 
       - \frac{1}{V^2}\left( v_at_b+ v_bt_a-g_{ab}v_ct^c \right)
\non \\
          &=& 0  \,,
\label{expr:Tab:s}
\eena
where we have used eq.~(\ref{formula:gab-ttepsepstt}) 
at the second line and eq.~(\ref{formula:eps-eps-t}) at the third line. 
This proves that we can express the gauge-invariant perturbation variable 
$Z_{ab}$ as 
\bena   
 Z_{ab} = \left( \OD_a\OD_b -\frac{1}{\ell^2} g_{ab}\right)\phi_S \,.  
\label{eq:Zab}
\eena 
Note that the residual freedom of $\phi_S$, which leaves 
$Z_{ab}$ unchanged, is 
\bena
  \phi_S \to \phi_S + \phi_0 \,, \quad 
  \left( \OD_a\OD_b -\frac{1}{\ell^2} g_{ab}\right)\phi_0 = 0 \,.       
\label{freedom:phi}
\eena 

Substituting the expression~(\ref{eq:Zab}) and 
eq.~(\ref{einstein:ai}) into the $(a,b)$-components 
of the linearised Einstein equations given in Ref.~\cite{KIS2000}, 
we obtain
\bena
 \left(\OD_a\OD_b -\frac{1}{\ell^2} g_{ab} \right)E(\phi_S) = 0 \,, 
\label{eq:E}
\eena 
where 
\bena
 E(\phi_S) \equiv r^2 
                  \left(
                       \OD^a\OD_a -n\frac{\OD^ar}{r}\OD_a 
                       - \frac{k_S^2-n}{r^2} + \frac{n-2}{\ell^2}
                  \right)\phi_S \,.
\label{def:E} 
\eena 
Viewing eq.~(\ref{eq:E}) as three differential equations for $E$, 
we find the general solution, 
\bena
  E_{sol} = c_1 r + V \left(c_2 \sin t + c_3 \cos t \right) \,, 
\label{Esol}
\eena 
where $c_1,c_2,c_3$ are arbitrary constants 
and $V  =(-t^a t_a)^{1/2} = \sqrt{1+r^2/\ell^2}$. 

On the other hand, since eq.~(\ref{freedom:phi}) 
is the same as eq.~(\ref{eq:E}), the residual freedom 
$\phi_0$ is also given by 
\bena  
  \phi_0 = c'_1 r + V \left(c'_2 \sin t + c'_3 \cos t \right) \,,   
\label{sol:phi0}
\eena 
with arbitrary constants $c'_1,c'_2,c'_3$. 
Using these explicit forms of $E_{sol}$ and $\phi_0$, 
we see that  
under the change~(\ref{freedom:phi}), 
$E(\phi_S)$ transforms as 
\bena
  E(\phi_S) &\rightarrow& E(\phi_S + \phi_0)
\non\\
   &=& E_{sol} -(k_S^2-n)\phi_0 + n\frac{r^2}{\ell^2}\phi_0 
                           - n r\left(1+\frac{r^2}{\ell^2}\right)
                                \partial_r \phi_0 
\non \\
   &=& E_{sol} -(k_S^2-n)\phi_0 -n c'_1 r 
\non \\
   &=& c''_1 r 
        + V \left( c''_2 \sin t + c''_3 \cos t \right) \,, 
\label{transf:Ephi}
\eena 
where $c''_1=c_1 -k_S^2c'_1$ $c''_2 = c_2 -(k_S^2-n)c'_2$ 
$c''_3 = c_3 -(k_S^2-n)c'_3$. Thus, setting these arbitrary constants 
$c''_1, c''_2, c''_3$ zero, we can make $E(\phi_S+\phi_0)$ vanish.  
Therefore, from eq.~(\ref{def:E}), we obtain 
\bena 
     \OD^a\OD_a \Phi_S
     - \left\{ 
              \frac{(n-2)(n-4)}{4}\frac{1}{\ell^2}  
             + \left[\frac{n(n-2)}{4}+l(l+n-1)\right]\f(1/r^2)
      \right\} \Phi_S =0 \,, 
\label{eq:covariant-master-S}    
\eena 
where $\Phi_S \equiv r^{-n/2}(\phi_S+\phi_0)$. 

Again, the equation for scalar-type gravitational 
perturbations takes the form~(\ref{eq:canonical})
with $\nu^2$ given by
\bena
  \nu^2 =  \frac{(n-2)(n-4)}{4} +\frac{1}{4} \,, 
\label{def:nu:scalar}
\eena  
and $\sigma^2$ again given by eq.~(\ref{def:sigma}).

Thus, we have reduced the problem of solving the gravitational
perturbation equations to solving the 3 scalar equations
(\ref{eq:covariant-master-T}), (\ref{eq:covariant-master-V}), and
(\ref{eq:covariant-master-S}) for $\Phi_T$, $\Phi_V$, and $\Phi_S$,
respectively. 
Note that the vector- and scalar-type gravitational perturbations
in $4$-dimensions (i.e., $n=2$), and scalar-type perturbations 
in $6$-dimensions (i.e., $n=4$) satisfy the same equation as 
a conformally invariant scalar field, 
while in all dimensions the tensor-type gravitational perturbations
satisfy the same equation as a minimally coupled massless scalar field. 

Finally, we note that the tensor spherical harmonics have
$(n-2)(n+1)/2$ independent components, the vector spherical harmonics
have $(n-1)$ independent components, and the scalar spherical
harmonics have $1$ independent component. Thus, the total number of
independent components is $(n+2)(n-1)/2$, which corresponds to the
number of dynamical degrees of freedom for gravitational radiation in
an $(n+2)$-dimensional spacetime.

\section{Dynamics in anti-de Sitter spacetime} 
\label{sect:3} 

We have seen that the field equations for 
scalar fields, electromagnetic fields and gravitational perturbations 
can be reduced to a wave equation of the form 
\bena 
 { \partial^2 \over \partial t^2}\Phi  
 = \left(
      \frac{ \partial^2}{ \partial x^2}
     - \frac{\nu^2-1/4}{\sin^2x} 
     -\frac{\sigma^2-1/4}{\cos^2x}
   \right) \Phi \,, 
\label{can2}
\eena 
which is defined on the interval $(0,\pi/2)$, where $x= \pi/2$ corresponds
to the origin of spherical coordinates and $x=0$ corresponds to infinity
in AdS spacetime.
In all cases, $\sigma$ takes the value (see eq.~(\ref{def:sigma}))
\bena
\sigma = l + \frac{n-1}{2} \,.   
\label{sig}
\eena
On the other hand, the value of $\nu^2$ depends upon the field and the type
(i.e., scalar, vector, or tensor) of perturbation. For scalar fields, $\nu^2$
can take any (positive or negative) value, depending on the value of the
effective mass, $m^2 \equiv m_0^2 + \xi R$, of the scalar field. For the 
electromagnetic and gravitational cases, the values of $\nu^2$ are given by
eqs.~(\ref{em:nu:vector}), (\ref{em:nu:scalar}), 
(\ref{def:nu:tensor}), (\ref{def:nu:vector}), and (\ref{def:nu:scalar}) above.
In this section, we will determine all of the possible boundary conditions
that can be imposed at infinity ($x=0$) so as to yield sensible dynamics
for this equation.

\subsection{The operator $A$ and its properties}

The general prescription of \cite{Wald1980} for defining dynamics (see
section 1 above) corresponds to viewing 
the differential operator 
\bena
 A \equiv - \frac{ d^2}{ d x^2}  
     + \frac{\nu^2-1/4}{\sin^2x} 
     +\frac{\sigma^2-1/4}{\cos^2x} \,, 
\label{def:A} 
\eena 
appearing in eq.~(\ref{can2})
as an operator on the Hilbert space ${\cal H}=L^2([0,\pi/2],dx)$. 
Indeed, taking account of the field 
redefinition~(\ref{redef}), it is easily seen that 
for each spherical harmonic,
the inner product for $\phi(t,x,z^i)$ 
in $L^2(\Sigma,V^{-1}d\Sigma)$ corresponds to the inner product
for $\Phi(t,x)$ defined on ${\cal H}$, i.e.,  
\bena
 \int_\Sigma \!\psi^* \phi \; V^{-1}d\Sigma 
                   = \int^{\pi/2}_0 \!\Psi^* \Phi \; dx \,.    
\label{def:inner-x} 
\eena 
It is easily seen that, when defined
on the natural domain $C^\infty_0(0,\pi/2)$, $A$ is a symmetric operator.
For later purposes, it will be useful to classify $A$ according to 
the possible values of $\nu^2$ as follows:

\medskip 
\noindent
{\bf Case (i)}: $\nu^2 \geqslant 1 $. This case occurs for scalar
fields whose effective mass squared satisfies $m^2 \geqslant 
-(n+3)(n-1)/4\ell^2$.  In particular, this case encompasses the
minimally coupled massless scalar field in all dimensions. This case
also encompasses the tensor-type of gravitational perturbations in 
all dimensions where they are defined (namely 
spacetime dimensions greater than 
or equal to $5$, i.e., $n\geqslant 3$), 
the vector-type of electromagnetic and
gravitational perturbations in spacetime dimensions greater than or
equal to $5$ (i.e., $n\geqslant 3$), and the scalar-type of
electromagnetic and gravitational perturbations in spacetime
dimensions greater than or equal to $7$ (i.e., $n\geqslant 5$).

\medskip 
\noindent
{\bf Case (ii)}: $0<\nu^2 <1$. This case occurs for scalar fields
whose effective mass is in the range $-(n+1)^2/4\ell^2 < m^2 <
-(n+3)(n-1)/4\ell^2$. (Note that in $4$ spacetime dimensions, this relation
becomes $-9/4\ell^2 < m^2 < -5/4\ell^2$, which corresponds to the case
analysed by Breitenlohner and Freedman~\cite{BF1,BF2}.) In particular,
this case encompasses the conformally invariant scalar field in all
spacetime dimensions. This case also encompasses the vector-type of
electromagnetic and gravitational perturbations in $4$-spacetime
dimensions (i.e., $n=2$), and the scalar-type of electromagnetic and
gravitational perturbations in $4$- and $6$-spacetime dimensions
(i.e., $n=2,\,4$).

\medskip 
\noindent
{\bf Case (iii)}: $\nu^2=0$. This case occurs for a scalar field whose
effective mass squared takes the critical value $m_C^2 \equiv
-(n+1)^2/4\ell^2$.  It also occurs for scalar-type of electromagnetic and
gravitational perturbations in $5$-dimensional (i.e., $n=3$) AdS
spacetime.

\medskip 
\noindent 
{\bf Case (iv)}: $\nu^2<0$. This case occurs only for a scalar field
whose effective mass satisfies
$m^2 <-(n+1)^2/4\ell^2$. 

\medskip

The positivity properties of $A$ are governed by the following proposition:

\medskip  
\noindent
{\bf Proposition 3.1}: If $\nu^2 \geqslant 0$ 
(i.e., in cases (i), (ii), and (iii) above), 
then $A$ is a positive operator. On the other hand, if $\nu^2 < 0$
(i.e., in case (iv) above) then $A$ is unbounded below.

\medskip 
\noindent 
{\bf Proof}: 
Consider, first, the case $\nu^2 \geqslant 0$. Let $\nu$ denote the positive
square root of $\nu^2$ and write
\bena         
  G_\nu(x) \equiv (\cos x)^{\sigma +1/2} \cdot (\sin x)^{-\nu +1/2} \,.  
\label{def:G} 
\eena
We define the operators ${\tilde D}^{\pm}$ by
\bena
  {\tilde D}^{\pm}\equiv \pm \frac{d}{d x} 
                    - {G_\nu}^{-1}\frac{d {G_\nu}}{dx}  \,,    
\eena 
Then $A$ can be expressed as
\bena
   A = {\tilde D}^{-} {\tilde D}^{+} + (-\nu + \sigma +1)^2 \,.   
\eena 
For arbitrary $\Phi \in {\rm Dom}(A) = C^\infty_0(0,\pi/2)$, it is
then easily seen that
\bena
 (\Phi, A\Phi)_{L^2} = (-\nu + \sigma +1)^2 \int_0^{\pi/2}|\Phi|^2dx 
           + \int_0^{\pi/2}|\tilde{D}^+ \Phi|^2dx \,, 
\label{eq:Apos}
\eena 
which establishes the positivity of $A$.

Next, we consider the case $\nu^2 < 0$.
Let $f$ be the $C^1$ function
\bena    
       f (x) \equiv \left\{ 
        \begin{array}{@{\,}ll} 
             (\sin x)^2 & \quad 0 < x \leqslant \eta 
\\
            (\cos x)^2 
             \left\{
                   p(\sin x)^{1/2}  + q (\sin x)^{-1/2} 
             \right\} 
              & \quad \eta < x \leqslant \pi/2     
        \end{array} 
                   \right.  \,,  
\eena 
where $0<\eta<\pi/2$ and 
\bena 
 p = \frac{(\sin \eta)^{3/2} \{4+(\cos \eta)^2 \} }{2(\cos \eta)^4} 
\,, \quad 
 q = \frac{(\sin \eta)^{5/2} \{-4 + (\cos \eta)^2 \} }{2(\cos \eta)^4} 
\,. 
\eena 
Straightforward calculation shows that for small $\eta$, 
\bena
&&{}
  \int ^{\pi/2}_0  f
                     \left( 
                          - \frac{\partial^2}{\partial x^2} 
                            + \frac{\nu^2-1/4}{\sin^2 x}  
                            + \frac{\sigma^2-1/4}{\cos^2 x}  
                     \right) fdx
  \Bigg/ \int ^{\pi/2}_0 |f|^2  dx 
 \non \\ 
   &&{} 
     = - 5\nu^2 \log \eta 
         + C + O(\eta) \,,      
\label{eq:logepsilon}
\eena
where $C$ is a constant independent of $\eta$.  
Hence, for $\nu^2 <0$, by taking $\eta \rightarrow 0$, the right hand side 
of eq.~(\ref{eq:logepsilon}) can be made to take arbitrarily large 
negative values. However, this does not yet prove the desired result, since
$f \notin {\rm Dom}(A) = C^\infty_0(0,\pi/2)$.
To remedy this, we smooth $f$ as follows. 
Let $\chi$ be an arbitrary smooth 
function with support contained within $(-\pi/4, \pi/4)$
satisfying $\int \chi(x) dx = 1$. Define
$\chi_\epsilon (x) = \epsilon^{-1} \chi(x/\epsilon)$ (so that 
$\chi_\epsilon$ approaches a $\delta$-function as 
$\epsilon \rightarrow 0$) and define
\bena
  f_\epsilon(x) \equiv \int \chi_\epsilon(x-x') 
  f(-\epsilon + (1+\epsilon)x') dx'  \,, 
\label{feps} 
\eena
where we have extended the domain of definition of $f$ beyond $[0,\pi/2]$
by setting $f(x) = 0$ whenever $x \notin [0,\pi/2]$.
Then, for sufficiently small $\epsilon$, we have $f_\epsilon 
\in C^\infty_0(0, \pi/2) = {\rm Dom}(A)$, so formula (\ref{eq:Apos}) holds.
However, since $f$ is $C^1$ and goes to zero sufficiently rapidly
at $x=0, \pi/2$,
it is straightforward to check that as $\epsilon \rightarrow 0$, 
$f_\epsilon$ and 
${\tilde D}^{+}f_\epsilon$ converge uniformly on $[0, \pi/2]$ to 
$f$ and ${\tilde D}^{+}f$,
respectively. It then follows that
$(f_\epsilon ,A f_\epsilon)_{L^2}/ \|f_\epsilon\|_{L^2}^2$ 
converges to the left side of
eq.~(\ref{eq:logepsilon}) as $\epsilon \rightarrow 0$. 
Consequently, there exist $\Psi \in {\rm Dom}(A)$ such that
$(\Psi,A \Psi)_{L^2}/ \|\Psi\|_{L^2}^2$ takes arbitrarily large negative
values.
\hfill $\Box$ 

\medskip

Thus, in case (iv), no positive, self-adjoint extensions of $A$ exist,
and there is no way of defining a sensible, stable dynamics. On the
other hand, at least one positive, self-adjoint extension (namely, the
Friedrichs extension) must exist in cases (i), (ii), and (iii).  Our
task is to find all of the positive, self-adjoint extensions of $A$ in
these cases.

There is a well known
systematic method, due to von Neumann, for obtaining 
all of the self-adjoint extensions of an arbitrary symmetric operator
on a Hilbert space ${\cal H}$. We will now review this method 
(see e.g., Ref.~\cite{RS1975} for proofs and further details).  
Let $A$ be a symmetric operator with domain ${\rm Dom}(A)$, 
and let $A^*$ denote the adjoint of $A$.  
Consider the subspaces $\K_\pm \subset {\cal H}$ (called 
{\em deficiency subspaces}) spanned by 
$\Phi_\pm \in {\rm Dom}(A^*)$ which satisfy respectively  
\bena
   A^* \Phi_\pm = \pm i \Phi_\pm \,.  
\label{eq:deficiency}
\eena 
Let $n_\pm = {\rm dim}({\K}_\pm)$. (Either or both of
$n_+$ and $n_-$ may be infinite.)
If $n_+ \neq n_-$, then no self-adjoint extension of $A$ exists.
If $n_+ =n_- = 0$, then there exists a unique self-adjoint extension, obtained
by taking the closure of $A$ (see e.g., Ref.~\cite{RS1975}). On the other
hand, if $n_+ =n_- \neq 0$, then all of the self-adjoint extensions of $A$
can be constructed as follows: Let $U: \K_+ \rightarrow \K_-$ be a
partial isometry, i.e., $\|\Phi_+\| =\|U \Phi_+\|$ for all
$\Phi_+ \in \K_+$. Given $U$, we define the associated self-adjoint
extension, $A_E$, of $A$ by first taking the closure, $\bar{A}$, of $A$ 
and then extending the 
domain of $\bar{A}$ by adding to it all elements of the form 
$\Phi_U = \Phi_+ + U \Phi_+$, i.e., we define 
\bena
{\rm Dom}(A_E) = \{ \Phi_0 + \Phi_+ + U \Phi_+ | \Phi_0 \in
{\rm Dom}(\bar{A}), \Phi_+ \in \K_+ \} \,. 
\label{domAE}
\eena
For all $\Phi \in {\rm Dom}(A_E)$, we define
\bena
A_E \Phi = \bar{A} \Phi_0 + i\Phi_+ - i U \Phi_+  \,. 
\label{AE}
\eena
Then $A_E$ is self-adjoint, and all self-adjoint extensions of $A$ can
be obtained in this manner. Thus, the self-adjoint extensions of $A$
are in one-to-one correspondence 
with the partial isometries $U: \K_+ \rightarrow \K_-$.

{}From the definition of $A^*$, it can be seen immediately that
eq.~(\ref{eq:deficiency}) is equivalent to the statement that
for all $\Psi \in {\rm Dom}(A)$, we have
\bena
 ((A \pm i)\Psi, \Phi_\pm)_{\cal H} = 0 \,.
\label{eq:deficiency2}
\eena
In our case, $A$ is an elliptic differential operator with domain
$C^\infty_0(0,\pi/2)$ on the Hilbert space ${\cal H}=L^2([0,\pi/2],dx)$.
It follows immediately that $\Phi_\pm$ satisfies the equation
$(A \mp i) \Phi_\pm = 0$ in the distributional sense. By elliptic regularity
(see subsection 2.1 above),
it follows that $\Phi_\pm$ must be smooth on $(0, \pi/2)$. Thus, in our case,
the solutions to eq.~(\ref{eq:deficiency}) are precisely the smooth functions
on $(0, \pi/2)$ that satisfy the ordinary differential equation 
\bena 
    \left( 
          - {d^2 \over dx^2} 
          + {\nu^2 - 1/4 \over \sin^2x} 
          + {\sigma^2 - 1/4 \over \cos^2x} 
    \right) \Phi_\pm
    = \pm i \Phi_\pm \,,  
\eena 
and are square integrable on $(0, \pi/2)$. We turn now to an analysis of
solutions to this equation.

\subsection{General solutions to the eigenvalue equation} 
In this subsection, we analyse solutions to the ordinary 
differential equation
\bena 
    \left( 
          - {d^2 \over dx^2} 
          + {\nu^2 - 1/4 \over \sin^2x} 
          + {\sigma^2 - 1/4 \over \cos^2x} 
    \right) \Phi
    = \omega^2 \Phi \,,  
\label{ode:formal}
\eena 
with $\sigma$ given by eq.~(\ref{sig}), with $\nu$ an 
arbitrary non-negative real number, $\nu \geqslant 0$. This encompasses
cases (i), (ii), and (iii) of the previous subsection. As discussed in the
previous subsection, the self-adjoint extensions of $A$ can be determined
by finding all the square integrable solutions 
to this equation for $\omega^2 = \pm i$.
More generally since the problem of finding the self-adjoint extensions of
$A$ is obviously equivalent to the problem of finding the self-adjoint 
extensions of $a A$ for any positive real number $a$,
the self-adjoint extensions of $A$ can be determined
by finding the square integrable solutions 
to eq.~(\ref{ode:formal}) for $\omega^2 = \pm a i$ for any
$a > 0$. It will turn out 
to be slightly more convenient to study this equation for $a = 2$, and
we shall do so below. However, we also will be interested later in
solutions to eq.~(\ref{ode:formal}) for $\omega^2 < 0$, i.e., $\omega$
imaginary. Therefore, at this stage we will allow
$\omega$ to be an arbitrary complex number.

The general solution to the equation~(\ref{ode:formal}) is given, 
in terms of hypergeometric functions, by 
\bena 
&& \Phi
  = (\sin x)^{\nu + 1/2} \cdot (\cos x)^{-\sigma + 1/2} 
  \non \\
 && {}\quad 
    \times 
     \left\{ 
           B_1 \cdot (\cos x)^{2\sigma} \cdot  
     F\left( 
            \zeta_{\nu,\sigma}^\omega, \zeta_{\nu,\sigma}^{-\omega},
            1+\sigma; \cos^2 x 
      \right)
          + {B}_2 \cdot \tilde{F} (\cos^2 x)\,\, 
     \right\} 
\label{sol:general}
\eena 
where $B_1$, $B_2 \in {\Bbb C}$, and here and hereafter we use 
the abbreviation 
\bena
  \zeta_{\nu, \sigma}^{\omega} 
   \equiv \frac{\nu + \sigma + 1 +\omega}{2} \,.  
\eena 
If $\sigma $ is not an integer (i.e., when the number of 
spacetime dimensions is even), the second solution,
$\tilde{F}(\cos^2 x)$, is given by
\bena 
 && {}
 \tilde{F}(\cos^2 x)
   = F \left(
             \zeta_{\nu,-\sigma}^\omega,\zeta_{\nu,- \sigma}^{-\omega}, 
            1-\sigma; \cos^2 x 
      \right) \,.  
\eena
If $\sigma +1$ is a natural number (i.e., when the number of 
spacetime dimensions is odd) then---with the exception of the case
(which we shall not consider further here)
where $\omega$ is real and either 
$0<\zeta_{\nu,\sigma}^{\omega} \leqslant \sigma$ 
or $0<\zeta_{\nu,\sigma}^{-\omega} \leqslant \sigma$---then 
we have~\cite{GR} 
\bena 
 && {}    
 \tilde{F}
  =  F \left(
             \zeta_{\nu,\sigma}^\omega, \zeta_{\nu,\sigma}^{-\omega}, 
             1+\sigma; \cos^2 x 
      \right) \cdot \log(\cos^2 x) 
 \non \\ 
  &&{} \, \qquad 
    + \sum_{k=1}^{\infty} 
      \frac{ \left(\zeta_{\nu,\sigma}^\omega \right)_{k} 
             \left(\zeta_{\nu,\sigma}^{-\omega} \right)_{k}
           }{\left(1 + \sigma \right)_{k} k! 
           }\cdot              
         \{ h(k)-h(0)\} 
            \cdot 
         (\cos x)^{2k} 
 \non \\
  &&{} \, \qquad 
    - \sum_{k=1}^{\sigma} 
      \frac{(k-1)!
             \left(-\sigma \right)_{k}
           }{ 
             \left(\zeta_{-\nu,-\sigma}^{\omega} \right)_{k}
             \left(\zeta_{-\nu,-\sigma}^{-\omega}\right)_{k} 
           } \cdot (\cos x)^{-2k} \,,           
\label{Ftil} 
\eena  
where  
\bena
 \left( \zeta \right)_{k} 
 &\equiv& \frac{\Gamma(\zeta+k)}{\Gamma(\zeta)} \,, 
\\
   \psi(z) &\equiv& \frac{d}{dz}\log \Gamma(z) \,,
\label{digamma} 
\\
   h(k) &\equiv& \psi(\zeta_{\nu,\sigma}^{\omega}+k) 
                + \psi(\zeta_{\nu,\sigma}^{-\omega} + k)
          - \psi(1+\sigma +k)- \psi(k+1) \,,   
\eena 
and for the $\sigma =0$ case, the third term in the right side of 
eq.~(\ref{Ftil}) should be ignored. 

\medskip 

We consider, now, the behaviour of these solutions near $x = \pi/2$, which
corresponds to the origin of spherical coordinates in AdS spacetime.
For $\sigma \geqslant 1$, it follows immediately from 
the general form~(\ref{sol:general}) of the solutions 
that the square integrability of $\Phi$ at $x=\pi/2$ 
requires $B_2=0$. However, the situation is more subtle for the cases
$\sigma = 1/2$ and $\sigma = 0$.
 
For the case $\sigma =1/2$---which occurs only for an 
S-wave ($l=0$) for a scalar field in $4$-dimensional 
(i.e., $n=2$) spacetime---the leading behaviour near $x = \pi/2$ is 
$\Phi(x) \approx B_2 = const$. Hence $\Phi$ is square-integrable
even when $B_2 \neq 0$, and it might appear that both solutions to
eq.~(\ref{ode:formal}) are acceptable. However, this is actually
an artifact of our having artificially removed the origin from the
spacetime when we used spherical coordinates
to separate variables. (A similar phenomenon occurs when analysing 
the spherically symmetric eigenstates of the Hydrogen atom in ordinary
quantum mechanics.) If we put the origin back and analyse $\Phi$
on the $3$-dimensional space $\Sigma$, we find that the solution
is not acceptable when $B_2 \neq 0$.
To see this explicitly, define a Cartesian coordinate system,  
$\{ X, Y, Z \}$, in a neighbourhood of the origin $x=\pi/2$ 
on a three-dimensional timeslice, $\Sigma$, 
such that $\pi/2-x = \sqrt{X^2+Y^2+Z^2}$. 
Then, from eq.~(\ref{redef}), 
the corresponding field, $\phi(X,Y,Z)$, defined on $\Sigma$ 
is expressed in this neighbourhood as 
\bena 
   \phi = \left(\frac{\sin x}{\cos x} \right)^{n/2} \Phi \scalar 
           \approx \frac{B_2 }{\sqrt{X^2+Y^2+Z^2}} \,.
\eena 
This is square integrable in a neighbourhood of the origin $X=Y=Z=0$ 
with the integration measure $V^{-1} d\Sigma = dXdYdZ$. 
However, let $\tilde{A}$ denote the $3$-dimensional operator on
$L^2(\Sigma,V^{-1} d\Sigma)$ corresponding to $A$.
the leading order behaviour of $\tilde{A} \phi$ near the origin is 
given by
\bena
 - VD^i(VD_i \phi) 
  &\approx& -(\partial^2/\partial X^2 
            + \partial^2/\partial Y^2 
            + \partial^2/\partial Z^2)\phi  
\non \\
  &=& 4\pi \cdot B_2 \cdot \delta(X)\delta(Y)\delta(Z) \,,  
\eena  
where $\delta(\cdot)$ is the Dirac delta function. As a consequence of
this additional $\delta$-function source,
the ``solution'', $\phi$, on $\Sigma$ when $B_2 \neq 0$, 
does not actually correspond to a distributional solution of 
$\tilde{A} \phi = \omega^2 \phi$ but rather to an equation that
contains an additional $\delta$-function source at the origin. In particular,
when $\omega^2 = \pm 2i$, we do not obtain a
solution to the analog of
eq.~(\ref{eq:deficiency}) on $\Sigma$. 
Therefore, only the solution 
with $B_2 =0$ is acceptable.

Similarly, for the case $\sigma=0$---which occurs only for the 
S-wave ($l=0$) 
for a scalar field in the $3$-dimensional (i.e., $n=1$) case---$\Phi$ 
behaves near $x=\pi/2$ as 
$\Phi \approx (\cos x)^{1/2} \left\{ B_2 \log (\cos^2 x) 
+ B_1 + \cdots \right\} $, which is 
square-integrable near $x=\pi/2$ even when $B_2 \neq 0$. 
However the original field $\phi$ defined on a $2$-dimensional 
surface $\Sigma$ behaves near the origin as 
\bena
  \phi \approx B_2 \log (X^2+Y^2) + B_1 + \cdots \,,
\eena 
with $\{X,Y \}$ being locally defined Cartesian coordinates as 
in the $\sigma = 1/2$ case. Again, if $B_2 \neq 0$,
then on the $2$-dimensional surface, $\Sigma$,
$\phi$ does not correspond to a solution to the
equation of interest
but rather to an equation with an additional $\delta$-function
source. Thus, again the solutions with $B_2 \neq 0$ are not acceptable.
Thus, hereafter we discard 
the second solution~$\tilde F(\cos^2 x)$ in all cases.  


Having set $B_2 = 0$, we 
now turn our attention to the boundary at $x=0$, corresponding to
infinity in AdS spacetime.  
In order to more easily see the asymptotic behaviour of $\Phi$ 
near the boundary, it is convenient to transform 
the argument of the hypergeometric function $F$ from $\cos^2 x$ to $\sin^2 x$. 
Using the linear transformation formulae for hypergeometric 
functions~\cite{Handbook1965}, we obtain the following expressions:

\noindent
For $\nu \neq 0$, $\nu \notin {\Bbb N}$, we have
\bena
&& \Phi= B_1 \cdot  
         (\sin x)^{\nu + 1/2} \cdot (\cos x)^{\sigma + 1/2} \cdot 
         F\left( 
                \zeta_{\nu,\sigma}^{\omega}, 
                \zeta_{\nu,\sigma}^{-\omega}, 
                1+\sigma; \cos^2 x 
          \right) 
\non \\
 && {}\,\,\,\,
       = B_1 \cdot G_\nu(x) \cdot 
        \Bigg\{ 
         \frac{\Gamma(1+\sigma) \Gamma(\nu)  
                }{
                  \Gamma(\zeta_{\nu,\sigma}^{\omega}) 
                  \Gamma(\zeta_{\nu,\sigma}^{-\omega})
                } \cdot 
           F\left(
                  \zeta_{-\nu,\sigma}^{\omega}, 
                  \zeta_{-\nu,\sigma}^{-\omega}, 
                  1-\nu; \sin^2 x 
           \right) 
 \non \\
 && \quad \quad \,\,  
      +    \frac{\Gamma(1+\sigma) \Gamma(- \nu) 
                }{ 
                  \Gamma(\zeta_{-\nu,\sigma}^{\omega}) 
                  \Gamma(\zeta_{-\nu,\sigma}^{-\omega})
                }\cdot 
           (\sin x)^{2\nu } \cdot 
           F\left(
                  \zeta_{\nu,\sigma}^{\omega}, 
                  \zeta_{\nu,\sigma}^{-\omega}, 
                  1+\nu; \sin^2 x 
            \right)        
      \Bigg\} \,,  
\label{transf:hypergeometric-nu}  
\eena
where $G_\nu(x)$ was defined in eq.~(\ref{def:G}) above.

\noindent
For $\nu \in {\Bbb N}$, we have 
\bena
\Phi &=& B_1 \cdot  
         (\sin x)^{\nu + 1/2} \cdot (\cos x)^{\sigma + 1/2} \cdot 
         F\left( 
                \zeta_{\nu,\sigma}^{\omega}, 
                \zeta_{\nu,\sigma}^{-\omega}, 
                1+\sigma; \cos^2 x 
          \right) 
\non \\ 
   &=& B_1 \cdot G_\nu(x) \cdot 
    \Bigg[\; 
      \frac{\Gamma(1+\sigma) \Gamma(\nu) 
          }{\Gamma(\zeta_{\nu,\sigma}^{\omega})
            \Gamma(\zeta_{\nu,\sigma}^{-\omega})
          }  
          \sum_{k=0}^{\nu -1} 
          \frac{(\zeta_{-\nu,\sigma}^{\omega})_k
                (\zeta_{-\nu,\sigma}^{-\omega})_k
               }{k! (1-\nu)_k } \cdot 
          (\sin x)^{2k} 
\non \\
  &&- \frac{(-1)^{\nu}\Gamma(1+\sigma) 
          }{\Gamma(\zeta_{-\nu,\sigma}^{\omega})
            \Gamma(\zeta_{-\nu,\sigma}^{-\omega})
          }  \cdot 
          (\sin x)^{2\nu} \cdot
          \sum_{k=0}^{\infty} 
          \frac{(\zeta_{\nu,\sigma}^{\omega})_k
                (\zeta_{\nu,\sigma}^{-\omega})_k
               }{k!(\nu+k)!} \cdot 
          (\sin x)^{2k} \cdot
          \Big\{ \log (\sin^2 x)  
\non \\
 &&  \qquad 
                 -\psi(k+1)-\psi(k+\nu+1) 
                 + \psi(\zeta_{\nu,\sigma}^{\omega} + k) 
                 + \psi(\zeta_{\nu, \sigma}^{-\omega} + k) 
          \Big\} \; 
     \Bigg] \,.   
\label{transf:hypergeometric-integer}  
\eena  

\noindent 
For $\nu =0$, we have 
\bena
 \Phi &=& B_1 \cdot (\sin x)^{1/2} \cdot (\cos x)^{\sigma +1/2} \cdot 
              F(\zeta_{0,\sigma}^{\omega}, \zeta_{0,\sigma}^{-\omega}, 
                1+\sigma ; \cos^2 x) 
\non \\
  &=& B_1 \cdot G_{0}(x)\cdot 
    \frac{\Gamma(1+\sigma)
          }{\Gamma(\zeta_{0,\sigma}^{\omega})
            \Gamma(\zeta_{0,\sigma}^{-\omega})
          }  
          \sum_{k=0}^{\infty} 
          \frac{(\zeta_{0,\sigma}^{\omega})_k
                (\zeta_{0,\sigma}^{-\omega})_k
               }{(k!)^2} \cdot 
          (\sin x)^{2k} 
\non \\
  && \qquad 
       \times 
          \left\{ 
                 2 \psi(k+1)-\psi(\zeta_{0,\sigma}^{\omega} + k) 
                 - \psi(\zeta_{0,\sigma}^{-\omega} + k) 
                 - \log (\sin^2 x) 
          \right\}\,.   
\label{transf:hypergeometric-0}  
\eena  
When $\nu \geqslant 1$ (i.e., case (i) above),  
we see from eq.~(\ref{transf:hypergeometric-nu}) for 
$\nu \notin {\Bbb N}$ 
(and eq.~(\ref{transf:hypergeometric-integer}) for $\nu \in {\Bbb N}$) 
that $\Phi$ fails to be square integrable, except for the case in which  
$\omega$ is a real number such that either 
$\zeta_{\nu,\sigma}^{\omega}$ or $\zeta_{\nu,\sigma}^{-\omega}$  
becomes zero or a negative integer, i.e., 
\bena
  \omega = \mp( 2m + 1 + \nu + \sigma) \,, \quad m = 0,\,1,\,2,\,\dots
  \,.   
\label{eigenmodes}
\eena 
(When eq.~(\ref{eigenmodes}) holds, then
either $\Gamma(\zeta_{\nu,\sigma}^{\omega})$ or 
$\Gamma(\zeta_{\nu,\sigma}^{-\omega})$ diverges, so
the first term of the right side of 
eq.~(\ref{transf:hypergeometric-nu}) for $\nu \notin {\Bbb N}$ 
(or the first term of eq.~(\ref{transf:hypergeometric-integer}) 
for $\nu \in {\Bbb N}$) vanishes.) 
When $0<\nu<1$ (case (ii)) 
we see from eq.~(\ref{transf:hypergeometric-nu}) that $\Phi$ is 
square integrable for all $\omega \in {\Bbb C}$.   
When $\nu = 0$ (case (iii))
we see from eq.~(\ref{transf:hypergeometric-0}) that $\Phi$ also is
square-integrable for all $\omega \in {\Bbb C}$.

\subsection{Self-adjoint extensions}
   
In this subsection, we shall determine all of the positive self-adjoint
extensions of the operator $A$, eq.~(\ref{def:A}) that satisfy the 
regularity condition at the
origin, $x=\pi/2$, discussed in the previous subsection.
(As discussed in the previous subsection, this additional 
regularity condition need be imposed only in the case of the
$l=0$ modes of a scalar field in $3$ and $4$ spacetime dimensions.)
We have already
seen in Proposition 3.1 that if $\nu^2 < 0$ (case (iv)), $A$ itself is
unbounded below, so, clearly, no positive self-adjoint extensions
exist. Thus, we need only consider cases (i), (ii), and (iii). In these
cases, $A$ is positive, so there always
exists at least one positive self-adjoint extension.

As explained in subsection 3.1 and at the beginning of subsection 3.2,
we will determine all of the possible self-adjoint extensions of $A$
by finding all of the
square-integrable solutions to eq.~(\ref{ode:formal}) with $\omega^2 = \pm 2i$.
We have just seen that in case (i) (i.e., $\nu^2 \geqslant 1$), there are
no such solutions, so the deficiency subspaces are trivial. Hence, there
is a unique self-adjoint extension of $A$, which is automatically positive.
Thus, dynamics is well defined in this case, without the need to specify 
any additional boundary conditions at infinity. One may interpret this
result as being a consequence of the fact that the effective 
potential term in $A$
is sufficiently repulsive near infinity that the fields are completely
reflected back without ever reaching the boundary. Hence, no boundary
conditions at infinity need be imposed.

On the other hand, in cases (ii) and (iii), we found in the 
previous subsection that there is precisely 
one linearly independent square integrable solution
to the eigenvector equation (\ref{ode:formal}) with $\omega^2 = \pm 2i$
that satisfies the regularity
condition at the origin. Consequently, each of the deficiency subspaces
$\K_\pm$ (for the operator $2A$) is one-dimensional, 
and is spanned by the vector $\Phi_\pm$ of 
eqs.~(\ref{transf:hypergeometric-nu}) (for case (ii)) and 
(\ref{transf:hypergeometric-0}) (for case (iii)), with 
$\omega = 1 \pm i$, respectively, and with $(B_1)_\pm$ 
chosen so that $\|\Phi_+\|_{L^2} = \|\Phi_-\|_{L^2} = 1$. 

Since $n_+ = n_- = 1$ the only (partial) isometries from $\K_+$ to $\K_-$
are given by
\bena
U_\alpha \Phi_+ = e^{i \alpha} \Phi_-
\label{Ualpha}
\eena
where $\alpha \in (-\pi, \pi]$. 
Hence, in cases (ii) and (iii)
there exists a one-parameter family of self-adjoint extensions, $A_\alpha$,
of $A$, (see eqs.~(\ref{domAE}) and (\ref{AE})) parametrised by 
$\alpha \in (-\pi, \pi]$.
Our remaining tasks are (1) to interpret the choice of 
extension, $A_\alpha$, in terms of a choice of boundary condition 
at infinity and (2) to determine for what values of $\alpha$ 
the extension $A_\alpha$ is positive. 

In order to investigate both of these issues, we need to know the
asymptotic behaviour of 
\bena
\Phi_\alpha \equiv \Phi_+ + e^{i \alpha} \Phi_-
\label{Phialpha}
\eena
near $x=0$. For $0<\nu<1$, we obtain
\bena 
  \Phi_\alpha &\propto&  G_\nu(x)\cdot 
    \left\{
          a_\nu 
        + b_\nu (\sin x)^{2\nu}
        + c_\nu (\sin x)^{2} 
        + \cdots 
    \right\} \,, 
\label{behave:phi1-0<nu<1}
\eena 
where $G_\nu(x)$ was defined in eq.~(\ref{def:G}) above.
Here the coefficients of the two leading terms are given by
\bena  
  a_\nu &=& 
      - \frac{ 2 e^{i\alpha/2} \Gamma(1+\sigma) \Gamma(\nu)
             }{\left|\zeta_{\nu,\sigma}^{-(1+i)} \right|
               \left|
                   \Gamma\left(\zeta_{\nu,\sigma}^{-(1+i)}\right)
               \right|^2 
             } \cdot 
            \sin \left(\frac{\alpha}{2}-\theta_{+\nu} \right) \,, 
\label{def:anu}
\\   
  b_\nu &=& 
      - \frac{ 2 e^{i\alpha/2} \Gamma(1+\sigma) \Gamma(-\nu)  
             }{\left|\zeta_{-\nu,\sigma}^{-(1+i)} \right|
               \left| 
                  \Gamma\left(\zeta_{-\nu,\sigma}^{-(1+i)}\right)
               \right|^2 
             } \cdot 
        \sin \left(\frac{\alpha}{2}-\theta_{-\nu} \right) 
          = a_{-\nu} \,, 
\label{def:bnu}
\eena 
where we have used the facts that
$\zeta_{\nu,\sigma}^{1+i} = \zeta_{\nu,\sigma}^{-(1+i)}{}^* +1$,  
$\zeta_{\nu,\sigma}^{1-i}=\zeta_{\nu,\sigma}^{-(1+i)}+1$, 
and $\zeta_{\nu,\sigma}^{-(1-i)} = \zeta_{\nu,\sigma}^{-(1+i)}{}^*$,  
and where we have defined $\theta_{\pm \nu} \in (-\pi/2, \pi/2]$ by 
\bena
 \sin \theta_{\pm \nu} 
 \equiv \frac{\pm \nu+\sigma}{\sqrt{1+(\pm \nu +\sigma)^2}} \,.
\eena 

Similarly, for $\nu=0$ we obtain 
\bena 
  \Phi_\alpha \propto G_{0}(x)\cdot 
           \Big\{ a_0 \log (\sin^2 x) 
                + b_0 
                + c_0 (\sin x)^{2}\log (\sin^2 x) 
                + \cdots  
           \Big\} \,.
\label{behave:phi1-nu0}
\eena 
Using the property of the digamma function, 
$\psi(z+1)=\psi(z)+1/z$ (see eq.~(\ref{digamma})), 
we find that the leading two coefficients are 
\bena 
&& a_0 = 
       \frac{ 2 e^{i\alpha/2} \Gamma(1+\sigma) 
            }{ 
            \left|\zeta_{0,\sigma}^{-(1+i)} \right|
            \left|\Gamma\left(
                              \zeta_{0,\sigma}^{-(1+i)} 
                        \right)
            \right|^2 
            } 
	 \cdot 
         \sin \left( \frac{\alpha}{2} - \theta_{0} \right) \,, 
\label{def:a0}
\\
&&b_0 = \frac{ 2 e^{i\alpha/2} \Gamma(1+\sigma)
              }{
              \left|\zeta_{0,\sigma}^{-(1+i)} \right|
              \left|
                  \Gamma\left(\zeta_{0,\sigma}^{-(1+i)}\right)
              \right|^2
              } \cdot
         \Bigg[
               2\cos^2 \theta_0 \cdot  
               \cos \left(
                          \frac{\alpha}{2} - \theta_0
                    \right)
\non \\ 
 && \qquad 
              + 2
               \left\{
                      \gamma
                     + \re \left[
                                 \psi\left(
                                          \zeta_{0,\sigma}^{-(1+i)}
                                     \right)  
                           \right]
                     + \sin \theta_0 \cos \theta_0
           \right\}\cdot
               \sin \left( \frac{\alpha}{2} - \theta_{0} \right)
         \Bigg] \,, 
\label{def:b0} 
\eena  
with $\gamma$ being Euler number 
and with $\theta_{0} \in [0, \pi/2)$ defined by 
\bena
  \sin \theta_{0} \equiv \frac{\sigma}{\sqrt{1+\sigma^2}} \,.
\eena 

By inspection of eqs.~(\ref{def:anu})-(\ref{def:bnu})
we see that in case (ii) (i.e., $0<\nu<1$),
the value of $\alpha$ is in one-to-one correspondence with the value of the
ratio $b_\nu/a_\nu$, which may take any real value (including $\pm \infty$, 
corresponding to $a_\nu = 0$). 
Thus, in case (ii), the choice of self-adjoint extension of $A$ is uniquely
specified by the value of $b_\nu/a_\nu$, and a 
self-adjoint extension exists for any specified real value of this quantity
(including $\pm \infty$). Similarly, inspection of
eqs.~(\ref{def:a0})-(\ref{def:b0}) shows that in case (iii)
(i.e., $\nu = 0$), the self-adjoint
extensions of $A$ are uniquely characterised by the ratio $b_0/a_0$, 
which may assume any real value (including $\pm \infty$).

The relationship between the choice of extension and the
choice of boundary conditions can now be seen as follows. It is
easily seen that for arbitrary initial data in $C_0^\infty$
the general evolution law (\ref{def:dynamics}) yields 
$\phi_t \in {\rm Dom} (A_E)$. However, from eq.~(\ref{domAE}), we see that
${\rm Dom} (A_E)$ consists of vectors in the domain of the closure of 
$A$---which have as rapid fall-off as is possible at the 
boundary---together with vectors of the form
$\Phi_\alpha = \Phi_+ + U \Phi_+$. Therefore, the boundary conditions 
that are satisfied by vectors of this latter
form are the boundary conditions that will be satisfied by all solutions
that arise from arbitrary initial data in $C_0^\infty$. Consequently, 
the choice
of self-adjoint extension in cases (ii) and (iii) corresponds precisely
to the specification of the allowed asymptotic behaviour 
(as determined by the values of $b_\nu/a_\nu$ and $b_0/a_0$, respectively)
of solutions near infinity.

The interpretation of the allowed
boundary conditions at infinity is most apparent in the conformally
invariant case, $\nu =1/2$,
since in this case, the effective potential term in the operator 
$A$ is regular at the boundary. For $\nu =1/2$, 
it is easily seen from eq.~(\ref{behave:phi1-0<nu<1})
that we have
\bena
\left.\frac{\partial \Phi_\alpha/\partial x}{\Phi_\alpha}\right|_{x=0} 
 = \frac{b_\nu}{a_\nu} \,. 
\eena 
Therefore, we see that the choice $b_\nu/a_\nu = \pm \infty$ 
(i.e., $a_\nu = 0$) corresponds to 
the Dirichlet boundary condition, $\Phi_\alpha|_{x=0} = 0$, and the choice 
$b_\nu/a_\nu = 0$ corresponds to the Neumann boundary condition,
$\partial \Phi_\alpha/ \partial x|_{x=0} = 0$.
The other real choices of $b_\nu/a_\nu$ correspond to general Robin 
boundary conditions, in which a linear combination of $\Phi_\alpha|_{x=0}$
and $\partial \Phi_\alpha/ \partial x|_{x=0}$ is required to vanish.

{}For $\nu \neq 1/2$, the effective potential term in $A$ 
is divergent at $x=0$ and the ratio 
$(\partial \Phi_\alpha/\partial x)/\Phi_\alpha$ itself is no longer 
well-defined at $x=0$. Nevertheless, for $0<\nu<1$, the dominant
behaviour of $G_\nu^{-1}\Phi_\alpha$ as $x \rightarrow 0$ is governed by
$a_\nu$, whereas the dominant behaviour of 
$\partial (G_\nu^{-1}\Phi_\alpha)/ \partial x$ is governed by $b_\nu$. 
Therefore, for $0<\nu<1$, it seems natural to define 
``generalized Dirichlet conditions'' as corresponding to the case
$a_\nu = 0$ and ``generalized Neumann conditions'' as corresponding
to the case $b_\nu = 0$. 
The other choices of self-adjoint extensions may be viewed as corresponding
to generalized Robin conditions\footnote{ 
Generalized Robin boundary conditions play a role in the AdS-CFT
correspondence when one considers the double-trace perturbations of 
the quantum field theory on the boundary at infinity. 
In this context, 
the ratio $b_\nu/a_\nu$ appears in the action for the boundary 
field theory as a coupling constant to the double-trace interaction,  
and generalized Dirichlet and Neumann conditions 
correspond to fixed points of a renormalization group flow 
in the quantum field theory on the boundary~\cite{Witten2001} 
(see also, e.g.,~\cite{GM2003,GK2003,NO2004,Minces2004} 
and references therein).}. 
The Friedrich's extension corresponds to 
the choice $a_\nu = 0$, i.e., to generalized Dirichlet conditions.

{}For $\nu = 0$, the dominant behaviour of both $G_0^{-1} \Phi_\alpha$ 
and $\partial
(G_0^{-1}\Phi_\alpha)/ \partial x$ as $x \rightarrow 0$ is governed by $a_0$.
Thus, we may interpret the self-adjoint extension with $a_0 = 0$ as 
corresponding to the simultaneous imposition of 
generalized Dirichlet and Neumann
conditions. This extension is precisely the Friedrich's extension.

We turn now to the determination of which of the self-adjoint extensions
$A_\alpha$ are positive.
Our main results are given by the following theorem:

\medskip  
\noindent
{\bf Theorem 3.2}: Let $A$ be the operator eq.~(\ref{def:A}) on the Hilbert
space $L^2([0,\pi/2])$, defined on the domain $C^\infty_0(0,\pi/2)$. Then
the self-adjoint extensions of $A$---satisfying the
regularity condition\footnote{As discussed in subsection 3.2, 
this regularity condition is relevant only for 
the $l=0$ modes of a scalar field in $3$ and $4$ 
spacetime dimensions.} at the origin $x=\pi/2$ (see subsection 3.2)---and 
the positivity properties of these extensions are as follows:

\noindent
{\bf case (i)} If $\nu^2 \geqslant 1 $, then there exists a unique 
self-adjoint extension, and it is positive. 

\noindent
{\bf case (ii)} If $0<\nu^2 <1$, then the self-adjoint
extensions of $A$ comprise a one-parameter family, $A_\alpha$.
These extensions are characterised by
the ratio $b_\nu/a_\nu$, which may assume any real value, 
but the extensions are positive if and only if $b_\nu/a_\nu$ satisfies
\bena    
\frac{b_\nu}{a_\nu} 
      \geqslant 
     - \left| \frac{\Gamma(-\nu)}{\Gamma(\nu)} \right| 
       \frac{\Gamma\left(\zeta^0_{\nu,\sigma}\right)^2
            }{\Gamma\left(\zeta^0_{-\nu,\sigma}\right)^2}  \,, 
\label{condi:ii}
\eena 
(This inequality includes the case $b_\nu/a_\nu = \infty$, i.e., 
generalized Dirichlet conditions, $a_\nu = 0$.) 
In terms of $\alpha$, this condition is expressed as 
\bena
  \frac{\sin \left( \alpha/2-\theta_{-\nu}\right)
        }{\sin \left( \alpha/2-\theta_{+\nu}\right)} 
  \leqslant 
  \sqrt{\frac{1+(-\nu+\sigma)^2}{1+(\nu + \sigma)^2}} \cdot 
  \frac{\left|\Gamma\left( \zeta^{-(1+i)}_{-\nu,\sigma}\right)\right|^2 
       }{ \left|\Gamma\left(\zeta^{-(1+i)}_{\nu,\sigma}\right)\right|^2 
        }
  \frac{ \Gamma\left(\zeta^0_{\nu,\sigma} \right)^2 
       }{\Gamma\left(\zeta^0_{-\nu,\sigma} \right)^2 
        } \,. 
\eena 

\noindent
{\bf case (iii)} If $\nu^2=0$, then 
the self-adjoint
extensions of $A$ comprise a one-parameter family, $A_\alpha$.
These extensions are characterised by
the ratio $b_0/a_0$, which may assume any real value, 
and are positive if and only if 
\bena
\frac{b_0}{a_0} \leqslant 2\gamma + 2\psi\left(\zeta^0_{0,\sigma}
                                         \right) \,.  
\label{condi:iii} 
\eena  
(This inequality includes the case $b_0/a_0 = -\infty$, 
i.e., generalized Dirichlet-Neumann conditions, $a_0 = 0$.)  
In terms of $\alpha$, this condition is expressed as 
\bena
  \frac{\cos \left(\alpha/2 - \theta_0 \right)
      }{\sin \left(\alpha/2 - \theta_0 \right)}
   \leqslant (1+\sigma^2)\cdot 
             \left\{
                  \psi\left(\frac{\sigma +1}{2}\right)
                 -\re \left[
                            \psi \left(\frac{\sigma - i}{2}\right)
              \right]
                 - \frac{\sigma}{1 + \sigma^2}
             \right\} \,.
\eena 

\noindent  
{\bf case (iv)} If $\nu^2<0$, then $A$ itself is unbounded below, so all
self-adjoint extensions of $A$ are unbounded below.  

\medskip 
\noindent 
{\bf Proof}: 
We have already proven the above claims for cases (i) and (iv), and 
we have also established that in cases (ii) and (iii), the self-adjoint
extensions of $A$ comprise a one-parameter family, $A_\alpha$. 
Therefore, it remains only to prove the positivity claims in 
cases (ii) and (iii).

It is worth noting that by a relatively straightforward calculation
$A_\alpha$ can be seen to be positive whenever $b_\nu/a_\nu \geqslant 0$
in case (ii) and when $a_0 = 0$ in case (iii).
Namely, to prove the positivity of $A_\alpha$, it suffices to prove
its positivity on the smaller domain ${\cal D}_\alpha$ defined by
\bena
{\cal D}_\alpha = \{ \Phi_0 + c \Phi_\alpha | \Phi_0 \in
{\rm Dom}(A), c \in {\Bbb C} \} \,,
\label{calD}
\eena
since it is clear that $A_\alpha$ is the closure of its restriction to
${\cal D}_\alpha$ and positivity is preserved under
taking closures. Note that all $\Phi \in {\cal D}_\alpha$ are smooth.
Furthermore, since $A_\alpha (\Phi_0 + c \Phi_\alpha) = A \Phi_0 + 
2i c \Phi_+ -2i c e^{i\alpha} \Phi_-$ (where the factor of $2$ 
difference from eq.~(\ref{AE}) arises from our present use of $\pm 2i$
rather than $\pm i$ in the definition of the deficiency subspaces), it follows 
that for all $\Phi \in {\cal D}_\alpha$, we have
\bena
 (\Phi, A\Phi)_{L^2} =
  \int ^{\pi/2}_0  \Phi^*
                     \left( 
                          - \frac{\partial^2}{\partial x^2} 
                            + \frac{\nu^2-1/4}{\sin^2 x}  
                            + \frac{\sigma^2-1/4}{\cos^2 x}  
                     \right) \Phi dx \,.
\label{AD}
\eena
As in Proposition 3.1, we may write
\bena
   A = {\tilde D}^{-} {\tilde D}^{+} + (-\nu + \sigma +1)^2 
\eena 
with
\bena
  {\tilde D}^{\pm}\equiv \pm \frac{d}{d x} 
                    - {G_\nu}^{-1}\frac{d {G_\nu}}{dx}  \,,    
\eena 
where $G_\nu(x)$ is given by eq.~(\ref{def:G}). 

In case (ii)---i.e.,
when $0<\nu<1$---it can be checked that ${\tilde D}^{+}\Phi$ is 
square-integrable for all $\Phi \in {\cal D}_\alpha$. Consequently,
integration by parts can be justified, and we obtain
\bena
 (\Phi, A\Phi)_{L^2} = \left[-\Phi^* \tilde{D}^+\Phi \right]^{\pi/2}_{0} 
      \!   + (-\nu + \sigma +1)^2 \int_0^{\pi/2} \! |\Phi|^2dx 
           + \int_0^{\pi/2} \! |\tilde{D}^+ \Phi|^2dx \,. 
\label{eq:phi-A-phi}
\eena 
The boundary term from the origin, $x=\pi/2$, vanishes. The boundary term 
from $x=0$ is easily evaluated to yield
\bena
 (\Phi, A_\alpha \Phi)_{L^2} 
 = 
   2\nu a_\nu^* b_\nu 
   + (-\nu +\sigma +1)^2 \|\Phi\|_{L^2}^2 
   + \|{\tilde D}^{+}\Phi\|_{L^2}^2 \,, 
\eena 
where $a_\nu$ and $b_\nu$ are given by eqs.~(\ref{def:anu}) and 
(\ref{def:bnu}). Note that $\nu a_\nu^* b_\nu$ is always real (since
$b_\nu/a_\nu$ is real for any self-adjoint extension), as must
be the case, since $A_\alpha$ is self-adjoint.
By inspection, $A_\alpha$ will be positive whenever
$a_\nu^* b_\nu \geqslant 0$, as is the case whenever 
$b_\nu/a_\nu \geqslant 0$, including the case $a_\nu = 0$.

In case (iii)---i.e., when $\nu = 0$---we observe  
from (\ref{behave:phi1-nu0}) that the asymptotic behaviour of 
$\tilde{D}^{+} \Phi_\alpha$ near $x=0$ is
\bena 
\tilde{D}^{+}\Phi_\alpha 
  =2 (\cos x)^{\sigma +3/2}
    \left\{
         a_0 (\sin x)^{-1/2}+c_0(\sin x)^{3/2} \log(\sin^2 x)+\cdots
    \right\} \,. 
\eena 
We see, therefore,
that $\tilde{D}^{+} \Phi_\alpha$ is {\it not} square integrable
except for the case when $a_0 = 0$. 
In this latter case, integration by parts is again justified, and we have 
\bena
 (\Phi, A_\alpha \Phi)_{L^2} 
    = (\sigma +1)^2 \|\Phi\|_{L^2}^2 + \|\tilde{D}^+ \Phi\|_{L^2}^2  
    \geqslant 0 \,,
\eena 
so $A_\alpha$ is positive, as claimed.

To determine the precise range of positivity of $A_\alpha$ in cases
(ii) and (iii), we note that 
since $A_\alpha$ is strictly positive on the original domain of $A$ 
(namely, the $C_0^\infty$ functions) and since the
deficiency subspaces are only $1$-dimensional, the negative spectral subspace
of $A_\alpha$ can be at most $1$-dimensional. If the 
strictly negative spectral subspace
of $A_\alpha$ is $0$-dimensional, then $A_\alpha$ is positive. On the other
hand, if the strictly negative spectral subspace
of $A_\alpha$ is $1$-dimensional, then $A_\alpha$ must have an eigenvector, 
$\Psi_\beta$, of negative eigenvalue, $-\beta^2$, with $\beta> 0$. 
By elliptic regularity, $\Psi_\beta$
must be a smooth solution to the differential equation~(\ref{ode:formal}),
\bena 
    \left( 
          - {d^2 \over dx^2} 
          + {\nu^2 - 1/4 \over \sin^2x} 
          + {\sigma^2 - 1/4 \over \cos^2x} 
    \right) \Psi_\beta
    = -\beta^2 \Psi_\beta \,.  
\label{Psibeta}
\eena
Furthermore, in order to lie in the closure of ${\cal D}_\alpha$, it 
is necessary that the asymptotic behaviour of $\Psi_\beta$ approach
that of $c \Phi_\alpha$ at $x=0$ for some constant $c \in {\Bbb C}$ in 
the sense that as $x \rightarrow 0$ we must have 
$\Psi_\beta - c \Phi_\alpha = o([\sin(x)]^{\nu + 1/2})$. 

In case (ii), it follows from our general solution, 
eq.~(\ref{transf:hypergeometric-nu}), with $\omega = i \beta$ 
that in order to have the required asymptotic behaviour of $\Psi_\beta$ 
as $x \rightarrow 0$, $\beta$ must satisfy
\bena
 \frac{ \left|\Gamma\left(\zeta_{\nu,\sigma}^{i\beta} \right)\right|^2
      }{ \left|\Gamma\left(\zeta_{-\nu,\sigma}^{i\beta}
                     \right)
         \right|^2 }
  = \frac{b_\nu \Gamma(\nu)}{a_\nu \Gamma(-\nu)} \,.
\label{condi:neg-eigenvalue}
\eena
It follows from the properties of the gamma-function that 
for fixed $\sigma \geqslant 0$ and fixed $\nu$ in the range $0<\nu<1$,
the left side of this equation 
can be made arbitrarily large by letting $\beta \rightarrow \infty$ and
takes its minimum at $\beta =0$. 
Taking into account the fact that
$\Gamma(\nu)/\Gamma(-\nu)$ is negative for $0<\nu<1$, it 
follows that a solution to eq.~(\ref{Psibeta}) with $\beta > 0$
satisfying the required asymptotic
behaviour exists if and only if 
\bena
\frac{b_\nu}{a_\nu} 
      <
     - \left| \frac{\Gamma(-\nu)}{\Gamma(\nu)} \right| 
       \frac{\Gamma\left(\zeta^0_{\nu,\sigma}\right)^2
            }{\Gamma\left(\zeta^0_{-\nu,\sigma}\right)^2} \,. 
\eena 
Thus, eigenvector of negative
eigenvalue exists---and hence $A_\alpha$ fails to be positive---if and only
if this equation holds. This proves the claims of the theorem 
for case~(ii).

Similarly, in case (iii), it follows from our general solution,
eq.~(\ref{transf:hypergeometric-0}),
that in order to have the required asymptotic behaviour 
as $x \rightarrow 0$, $\beta$ must satisfy
\bena
 2\gamma +\psi\left(\zeta_{0,\sigma}^{i\beta}\right) 
                     +\psi\left(\zeta_{0,\sigma}^{-i\beta}\right) 
               =  \frac{b_0}{a_0} \,.  
\eena
{}From the properties of the digamma function, $\psi(x)$, we see that the
left side achieves its minimum value when $\beta = 0$ and becomes arbitrarily
large when $\beta \rightarrow \infty$. Thus, we can solve this 
equation---and thereby obtain an eigenvector of $A_\alpha$ of 
negative eigenvalue---if and only 
\bena
\frac{b_0}{a_0} > 2\gamma + 2 \psi\left(\zeta_{0,\sigma}^0\right) \,. 
\eena 
This proves the claims of the theorem for case (iii).
\hfill $\Box$

\medskip

We conclude with several remarks.

%
%

As already noted, case (ii) was considered by
Breitenlohner and Freedman~\cite{BF1,BF2} for scalar fields in 4-dimensional 
AdS spacetime. They showed that
with the imposition of generalized Dirichlet or Neumann boundary conditions,
the initial value problem 
is well-posed in this case. 
Their choice of these boundary conditions 
arose from the requirement that an energy 
functional for the scalar field be finite, positive, 
and conserved. However, 
they considered only two types of energy functionals: 
the usual energy functional of a minimally coupled scalar field, and
a conformal (``improved'') energy functional.
The requirement that the usual energy functional be conserved
allows only the generalized Dirichlet boundary condition,
whereas conservation of the
conformal energy allows us to take either the generalized Dirichlet or 
the generalized Neumann boundary condition.
We wish to emphasise here that
for {\it any} choice of a positive self-adjoint extension $A_E$, 
it is always possible to define a conserved, 
positive energy $E$ for all $\Phi \in {\rm Dom}(A_E)$ by 
\bena
 E(\Phi) \equiv (\dot \Phi, \dot\Phi)_{L^2} 
         + (\Phi,A_E \Phi)_{L^2} \,,       
\label{def:energy} 
\eena
where $\dot\Phi \equiv \partial \Phi/\partial t$. The generalized 
Dirichlet and Neumann conditions considered in \cite{BF1,BF2}, are 
merely two values of the one-parameter family of choices
given by eq.~(\ref{condi:ii}). For each boundary condition in this
range, there is a corresponding conserved, positive 
energy (see also \cite{MR2001} for another definition of such 
a positive, conserved energy in the scalar field case). 

As we have proven above, in cases (ii) and (iii), $A$ also admits 
non-positive self-adjoint extensions; in case (iv), all self-adjoint
extensions are unbounded below. If one chooses a non-positive self-adjoint
extension, then one can still define dynamics in a manner analogous
to eq.~(\ref{def:dynamics}), but the dynamics will be unstable in the
sense that generic solutions will grow unboundedly in time (see \cite{w1991}).
Indeed, in case (iv), since $A$ is unbounded below,
generic solutions will always grow in time more rapidly
than $\exp(at)$ for any positive $a$. 

We note that our analysis shows that in AdS spacetime 
of any dimension, it is always possible to define a stable dynamics for
electromagnetic and gravitational perturbations. In spacetime dimension
$\geqslant 7$, the dynamics is unique, and no choice of boundary conditions
at infinity need be made at all. However, in $4$ and $6$ dimensions, 
one must make
a choice of boundary conditions (in the range defined by 
eq.~(\ref{condi:ii})) for the scalar type of electromagnetic 
and gravitational perturbations; 
in 4 dimensions, one must also choose boundary conditions 
in this range for the
vector type of electromagnetic and gravitational perturbations.
Similarly, in 5-dimensional AdS spacetime one must make a choice
of boundary conditions (in the range defined by eq.~(\ref{condi:iii})) 
for the scalar type of
electromagnetic and gravitational perturbations. Each of these choices
of boundary conditions corresponds to a different theory of classical
bulk dynamics.



Particularly interesting in the context of the AdS-CFT correspondence 
is the case in which the bulk geometry contains a black
hole~\cite{W1998}. In the Schwarzschild-AdS background, we also can 
obtain master equations for perturbations~\cite{KI2003a} analogous 
to those derived in this paper. We then expect the same 
classification (i)-(iv) of the corresponding operator $A$ to hold, 
because the classification (i)-(iv) is made by inspecting merely 
the asymptotic behaviour of perturbations near the boundary. 
However, since the effective potential term in the corresponding operator 
$A$ will change significantly near the black hole 
event horizon~\cite{IK2003}, we would expect that the precise 
relation between the positivity of self-adjoint extensions 
and the boundary conditions at infinity will change.

\bigskip 
\begin{center}
{\bf Acknowledgements}
\end{center}
We wish to thank Gary Gibbons, Sean Hartnoll, Stefan Hollands, 
and Hideo Kodama for many helpful discussions and useful suggestions. 
RMW wishes to thank the Yukawa Institute for its hospitality during 
the time some of this research was carried out, and AI wishes to thank 
the Enrico Fermi Institute for its hospitality during the time other 
parts of this research were carried out. This research was supported 
in part by the Japan Society for the Promotion of Science 
and by NSF grant PHY 00-90138 to the University of Chicago.

\bigskip 

\noindent 
\appendix 
\section*{Appendix} 
\label{sect:appendix}

We show here that for the $l=1$ modes, there exist no dynamical degrees 
of freedom for gravitational perturbations (i.e., no dipole 
gravitational radiation). (See also \cite{KIS2000}.)  

Consider, first, the vector type of perturbations.
For $l=1$ ($k_V^2=n-1$), $\vector_i$ becomes a Killing vector field 
on the $n$-sphere. Then $2D_{(i}h^{(1)}_T{}_{j)}$ in 
(\ref{decompose:hij}) vanishes, and therefore $H^{(1)}_T$ is not defined. 
In this case, we can find a gauge-invariant variable 
\bena
 Z^{(1)}{}_{ab} = r^2\OD_a\left(\frac{H^{(1)}{}_b}{r^2}\right)
                 - r^2\OD_b\left(\frac{H^{(1)}{}_a}{r^2}\right) \,,  
\eena
which has only a single independent component. 
Then the linearised Einstein equations for $Z^{(1)}{}_{ab}$ 
become 
\bena
  \OD^a(r^{n}Z^{(1)}{}_{ab}) = 0 \,, 
\eena 
and can be explicitly solved, yielding 
\bena 
 Z^{(1)}{}_{ab} = \epsilon_{ab} \frac{C}{r^{n}}  \,,  
\label{l=1mode:odd}
\eena 
where $C$ is a constant. Therefore the vector-type perturbation 
of the $l=1$ mode is a non-dynamical (i.e., time independent)
perturbation, which can be interpreted as describing a change
in the angular momentum of the background.  

\bigskip 

For $l=1$ ($k_S^2=n$) of scalar-type perturbation, 
$\BD_i\scalar$ is a conformal Killing vector field, 
that is, $\BD_i\BD_j\scalar + \gamma_{ij}\scalar$ identically vanishes. 
Hence the term $h^{(0)}_T$ in (\ref{decompose:hij}) is not present. 
Setting $H^{(0)}_T=0$ in the definitions (\ref{def:Z}) 
and (\ref{def:Zab}) (and $X_a \equiv -H^{(0)}{}_a$), 
we can still use $Z$ and $Z_{ab}$ as metric perturbation variables, 
but in this case, $Z$ and $Z_{ab}$ are no longer gauge-invariant 
and equation~(\ref{einstein:ai}) is not obtained. 
Under gauge-transformations generated by 
$\xi^{(0)}{}_a \scalar dy^a + \xi^{(0)} {\BD}_i \scalar dz^i $,  
we find in particular, $  Z^a{}_a - Z \rightarrow 
 Z^a{}_a - Z + \bar{\delta} (Z^a{}_a - Z)$ with 
\bena
  \bar{\delta}(Z^a{}_a - Z) 
  \equiv  
     -r^{n/2} \left\{  
                   \OD^c\OD_c -\frac{(n-2)(n-4)}{4} \frac{1}{r^2} 
                  -\frac{n(n+2)}{4} \frac{1}{\ell^2} 
             \right\} \Xi \,,  
\eena 
where $ \Xi\equiv-2r^{(n-4)/2}\xi^{(0)}$. 
This implies that we can make eq.~(\ref{einstein:ai}) hold 
by choosing $\Xi$ suitably, or in other words we can impose  
eq.~(\ref{einstein:ai}) as a gauge condition. 
In this way, even for the $l=1$ modes, we can recover the same 
expression of the linearised Einstein equations for 
$Z$ and $Z_{ab}$ as in the $l\geqslant 2$ case.  

We should note however that the gauge condition $Z^a{}_a - Z =0$ 
does not completely fix gauges; it is possible to take 
further gauge transformations generated by $\Xi$ that satisfies 
the homogeneous equation 
\bena 
 \left\{
        \OD^a\OD_a -\frac{(n-2)(n-4)}{4} \frac{1}{r^2} 
               -\frac{n(n+2)}{4} \frac{1}{\ell^2} 
 \right\} \Xi = 0 \,.    
\label{eq:residual-gauge}
\eena 
Therefore the remaining gauge degrees of freedom are characterised 
by two arbitrary functions of $t$, i.e., initial data 
$(\Xi, \partial \Xi/\partial r)$ on a Cauchy surface $r=r_0$ 
in the globally hyperbolic spacetime $(O^2,-g_{ab})$. 
 
Now we shall show that the solutions $Z$ and $Z_{ab}$ 
to the linearised Einstein equations are gauge modes; 
$Z$ and $Z_{ab}$ can be made to vanish by further gauge 
transformations.      
First, with the gauge condition~(\ref{einstein:ai}), 
repeating the same argument as in the $l\geqslant 2$ case 
(below eq.~(\ref{Zaa})), we can express $Z_{ab}$ 
in terms of a scalar function $\phi_\xi$ as 
\bena
 Z_{ab} = \left( \OD_a\OD_b -\frac{1}{\ell^2} g_{ab}\right)\phi_\xi \,,   
\eena 
where $\phi_\xi$ satisfies the linearised Einstein equation 
(see eqs.~(\ref{def:E}) and (\ref{Esol}) with $k_S^2=n$) 
\bena
 E(\phi_\xi)
   \equiv 
   r^2 \left(
        \OD^a\OD_a -n\frac{\OD^ar}{r}\OD_a 
        + \frac{n-2}{\ell^2}
  \right)\phi_\xi   
   = V \left( c_2 \sin t + c_3 \cos t \right) \,, 
\label{eq:Ein:l=1}  
\eena 
with $c_2, \, c_3$ being arbitrary constants. 
Note that the term $c''_1 r$, which may exist in the right hand side (see 
eq.~(\ref{transf:Ephi})), has been already set to zero 
by using the residual freedom of $\phi_\xi$ defined 
by eq.~(\ref{freedom:phi}). 
%
Next, we find the gauge transformation 
$Z_{ab} \rightarrow Z_{ab}'=  Z_{ab} + \bar{\delta} Z_{ab} $ with
\bena
 \bar{\delta}Z_{ab} 
 &\equiv& 
     - r^{n/2}\Biggl\{ 
                 \OD_a\OD_b \Xi 
                -\frac{(n-2)}{2}\left(
                                      \frac{\OD_ar}{r}\OD_b \Xi 
                                     + \frac{\OD_br}{r}\OD_a \Xi 
                                \right)
\non \\ 
  && \qquad  
            + (n-1)g_{ab} \frac{\OD^cr}{r}\OD_c \Xi 
            + \frac{n(n-2)}{4}\frac{(\OD_ar)\OD_br}{r^2}\Xi                 
            - \frac{n}{2}\frac{\OD_a\OD_br}{r}\Xi                 
\non \\  
  && \qquad  
            + (n-1)g_{ab} \left[  
                              \frac{1}{r^2}
                              - \frac{n}{2}\frac{(\OD^cr)\OD_cr}{r^2} 
                          \right] \Xi                          
         \Biggr\} \,,    
\eena 
where $\Xi$ satisfies eq.~(\ref{eq:residual-gauge}). 
To achieve the gauge $Z_{ab}'=0$, we consider on an initial Cauchy 
surface $r=r_0$ the equations    
\bena
 Z_{ab}' = \bar{\delta}Z_{ab}(\Xi)
 + \left(\OD_a\OD_b-\frac{1}{\ell^2}g_{ab}\right)\phi_\xi = 0 \,.     
\label{eq:gauge-equiv-class}
\eena 
{}The trace part of eq.~(\ref{eq:gauge-equiv-class}) is expressed as 
\bena 
&& Vr^{(n-2)/2}\frac{\partial }{\partial r} 
     \left(
            \frac{1}{Vr^{(n-2)/2}}\Xi 
     \right) 
 - \frac{V}{r^{n/2}} \frac{\partial}{\partial r} 
        \left( 
               \frac{r^{n/2}}{V} \Phi_\xi 
        \right) 
\non\\
&& \qquad \qquad 
   =  \frac{1}{nr^{(n+2)/2}V} 
         \left(c_2 \sin t  + c_3 \cos t \right) \,,  
\label{eq:gec:trace}
\eena 
where $\Phi_\xi \equiv r^{-n/2}\phi_\xi$ and we have used 
eqs.~(\ref{eq:residual-gauge}) and (\ref{eq:Ein:l=1}). 
Note that the constants $c_2$ and $c_3$ in the right side of 
this equation are actually required to 
vanish due to the consistency with the $(t,r)$-component 
of eq.~(\ref{eq:gauge-equiv-class}), and that in the case 
$\Phi_\xi$ obeys the homogeneous master equation $E(r^{n/2}\Phi_\xi)=0$.   
The $(t,t)$-component of eq.~(\ref{eq:gauge-equiv-class}) is 
expressed as 
\bena 
&&  \left\{
          \frac{1}{2} \frac{\partial V^2}{\partial r}  
          + (n-1)\frac{V^2}{r} 
    \right\} \frac{\partial }{\partial r} \Xi 
 \non \\
 && \qquad 
  - \frac{1}{V^2}\frac{\partial^2}{\partial t^2} \Xi 
  - \left\{
          \frac{n}{4r} \frac{\partial V^2}{\partial r}  
          - \frac{(n-1)}{r^2} 
          + \frac{n(n-1)}{2} \frac{V^2}{r^2} 
    \right\} \Xi 
\non \\
&& \qquad \qquad 
   = \left\{
            -\frac{1}{V^2}\frac{\partial^2}{\partial t^2} 
            + \frac{1}{2}\frac{\partial V^2}{\partial r}                
                         \frac{\partial }{\partial r}          
            + \left(\frac{n}{4r}\frac{\partial V^2}{\partial r} 
                    -\frac{1}{\ell^2}
              \right) 
     \right\} \Phi_\xi \,. 
\label{eq:gec:tt}
\eena 
For any $(\Phi_\xi, \partial \Phi_\xi/\partial r)$, 
solving eqs.~(\ref{eq:gec:trace}) and (\ref{eq:gec:tt}), we obtain 
the initial data $(\Xi, \partial \Xi /\partial r)$ that satisfy 
eq.~(\ref{eq:gauge-equiv-class}) on the initial surface $r=r_0$.    
Then we define $\Xi$ to be the solution of 
eq.~(\ref{eq:residual-gauge}) with these initial data so that 
$Z_{ab}'=0$ is achieved throughout $(O^2,-g_{ab})$.

\end{document}